\documentclass{article}
\usepackage{fullpage}
\usepackage[utf8]{inputenc}
\usepackage[T1]{fontenc}    
\usepackage{hyperref}       
\usepackage{booktabs}       
\usepackage{amsfonts}       
\usepackage{nicefrac}       
\usepackage{enumitem}
\usepackage{amsmath}
\usepackage{amssymb}
\usepackage{amsthm}
\usepackage{xcolor}
\usepackage{natbib}
\usepackage{graphicx}
\usepackage{subcaption}
\usepackage{multirow}
\usepackage{authblk}
\usepackage[bottom]{footmisc}

\newcommand{\ICML}{ICML}
\newcommand{\ICMLyear}{ICML 2020}

\newcommand{\commentout}[1]{}
\newcommand{\exprev}{{\sc experimental}}
\newcommand{\currev}{{\sc curated}}
\newcommand{\snrev}{{\sc self-nominated}}
\newcommand{\elrev}{{\sc colleague}}

\title{A Novice-Reviewer Experiment to\\ Address Scarcity of Qualified Reviewers in Large Conferences}

\author[${}^\spadesuit$]{Ivan Stelmakh}
\author[${}^\spadesuit$]{Nihar B. Shah}
\author[${}^\spadesuit$]{Aarti Singh}
\author[${}^{\heartsuit \diamondsuit}$]{Hal Daum{\'e} III}

\affil[${}^\spadesuit$]{School of Computer Science, Carnegie Mellon University}
\affil[${}^{\heartsuit}$]{University of Maryland, College Park}
\affil[${}^{\diamondsuit}$]{Microsoft Research, New York}

\date{\vspace{-13pt} \texttt{\{stiv,nihars,aarti\}@cs.cmu.edu, me@hal3.name} }

\begin{document}

\maketitle

\begin{abstract}
Conference peer review constitutes a human-computation process whose importance cannot be overstated: not only it identifies the best submissions for acceptance, but, ultimately, it impacts the future of the whole research area by promoting some ideas and restraining others. A surge in the number of submissions received by leading AI conferences has challenged the sustainability of the review process by increasing the burden on the pool of qualified reviewers which is growing at a much slower rate. In this work, we consider the problem of reviewer recruiting with a focus on the scarcity of qualified reviewers in large conferences. Specifically, we design a procedure for (i) recruiting reviewers from the population not typically covered by major conferences and (ii) guiding them through the reviewing pipeline. In conjunction with \ICMLyear{} --- a large, top-tier machine learning conference --- we recruit a small set of reviewers through our procedure and compare their performance with the general population of \ICML{} reviewers. Our experiment reveals that a combination of the recruiting and guiding mechanisms allows for a principled enhancement of the reviewer pool and results in reviews of superior quality compared to the conventional pool of reviews as evaluated by senior members of the program committee (meta-reviewers).
\end{abstract}

\section{Introduction}

Over the last few years, Machine Learning (ML) and Artificial Intelligence (AI) conferences have been experiencing rapid growth in the number of submissions: for example, the number of submissions to AAAI and NeurIPS --- popular AI and ML conferences --- more than quadrupled in the last five years. The explosion in the number of submissions has challenged the sustainability of the peer-review process as the number of \emph{qualified} reviewers is growing at a much slower rate~\citep{sculley19tragedy,shah19principled}. While especially prominent in ML and AI, the problem is present in many other fields where ``\emph{submissions are up, reviewers are overtaxed, and authors are lodging complaint after complaint}''~\citep{mccook2006peer}.

The disparity between growth rates of the submission and reviewer pools increases the burden on the reviewers, thereby putting a severe strain on the review process. According to the president of the International Conference on Machine Learning (ICML) board John Langford~\citep{langford18bubble}, 
\emph{``There is significant evidence that the process of reviewing papers in machine learning is creaking under several years of exponentiating growth.''} Hence, it is important to increase the number of qualified reviewers in the system to keep up with the growing number of submissions.

When the size of a conference is small, program chairs can extend the pool of reviewers by manually selecting new reviewers among researchers who have enough expertise in the area. The selection can be guided by the program chairs' understanding of who might be a good reviewer or by personal recommendations made by other senior members of the program committee. In what follows, we refer to the pool of reviewers manually constructed by the program chairs as the curated pool. However, with a massive increase in the scale of the conference, such a manual addition to the curated pool does not allow bringing in enough reviewers to cover the demand of the conference. The program chairs must then rely on alternative ways of reviewer recruiting.

With this motivation in mind, in the present paper we aim to design and evaluate modifications to the reviewer recruiting process that simultaneously address two challenges:
\begin{itemize}[itemsep=1pt, leftmargin=15pt, topsep=4pt]
    \item \textbf{Challenge 1.} To avoid overloading reviewers, conferences need to find new sources of reviewers as there are not enough curated reviewers to review all papers. 
    
    \item \textbf{Challenge 2.} Conferences need to ensure that newly added reviewers do not compromise the quality of the process, that is, are able to write reviews of quality at least comparable to the curated reviewer pool.
\end{itemize}

In the past, conference organizers have been trying to expand the reviewer pool by relaxing the qualification bar, that is, by allowing researchers who meet some minimal requirements such as having one or two relevant publications to join the pool of reviewers without further screening. For example, 1176 out of 3242 (that is, 36\%) of the reviewers in the NeurIPS 2016 conference were recruited by requesting authors of each submission to name at least one author who is willing to become a reviewer, and 70\% of these reviewers were PhD students: researchers at very early stages of their careers~\citep{shah2017design}. Such practices have now become conventional and are adopted by many other conferences, including a flagship conference in artificial intelligence AAAI that in 2020 invited self-nominated individuals with publication history in top venues, and in 2021 requires authors of submissions to be willing to become reviewers on request. Similarly, \ICML{} 2020 --- a flagship ML conference --- distributed a public call for reviewers and accepted self-nominated individuals with publication and reviewing history in top venues.

While the aforementioned innovations allow to enlarge the reviewer pool, little scientific evidence exists on the quality of reviews written by reviewers recruited through these novel procedures. NeurIPS 2016 compared the reviews written by curated reviewers with reviews sourced from authors of submissions in terms of numeric scores (overall score and several criteria scores) and inter-reviewer agreement~\citep{shah2017design}. The analysis did not reveal a significant difference between populations, only showing that author-sourced reviews were slightly harsher in scoring the clarity of submissions. However, we note that this analysis only operates with scores given by reviewers and does not address the \emph{quality} of reviews  --- perhaps the most important metric for success of the conference peer-review process --- which is largely determined by the textual part of the review. Other works provide anecdotal and empirical evidence that junior reviewers are more critical than their senior counterparts~\citep{mogul13hypercriticality,toor09hypercticality,tomiyama07hypercticality} and that ``\emph{graduate students seem to be unable to provide very useful comments}''~\citep{patat19dpr}. Thus, while the methods employed by leading conferences address the first challenge, it remains unclear if and how they address the second challenge of high quality reviewing.\footnote{After the experiment described in this paper was completed, the NeurIPS 2020 conference released an analysis of their review process~\citep{lin20data} which compared quality of reviews sourced from authors of submissions and reviews written by the curated pool of reviewers. In Section~\ref{section:discussion} we compare the results presented therein with those of the present study.}

In this work, in conjunction with the review process of \ICMLyear{} we conduct a threefold experiment: 
\begin{itemize}[itemsep=1pt, leftmargin=15pt, topsep=2pt]
    \item First, we recruit reviewers from the population not typically covered by the reviewer-selection process of major conferences. In that, we target the population of very junior researchers with limited or no publication/reviewing history most of whom do not pass the recruiting filters of \ICML{}. Conceptually, in contrast to the standard approach of selecting reviewers based on some proxy towards reviewing ability (e.g., prior publication and reviewing history), we evaluate candidates' abilities to review in an auxiliary peer-review process organized for the experiment. 
    
    \item Second, we add a select set of reviewers recruited through our experiment to the reviewer pool of the \ICML{} conference and guide them through the peer-review process by offering mentoring.
    
    \item Finally, we evaluate the performance of these novice reviewers by comparing them with the general population of the \ICML{} reviewer pool on multiple aspects. In doing so, we augment the past analysis of~\citet{shah2017design} by using an explicit measure of review quality (evaluated by meta-reviewers) in addition to indirect proxies.
\end{itemize}

An important aspect of our experiment is that most of the reviewers brought to the reviewer pool through our experiment would not have been considered in standard ways of recruiting. Hence, our experiment offers a principled way to enlarge the reviewer pool. As a by-product, the new pool of reviewers contributes to diversity of peer review resonating with the virtues such as increased scrutiny and variety of opinions outlined by~\citet{garisto19diversity}.  Moreover, we offer the new reviewers a more guided introduction to the reviewing process which is known to help novice reviewers to write better reviews~\citep{patat19dpr} and improve their own writing skills~\citep{kerzendorf2020distributed}. From the perspective of training reviewers, our experiment is conceptually similar to the initiative of \emph{Journal of Neuroscience}~\citep{picciotto18mentoring} and SIGCOMM conference~\citep{feldmann2005experiences} that attempt to help novices in becoming reviewers.

This work falls in the line of empirical works that study various behavioral aspects of human computation, including motivational aspect~\citep{kaufmann11fun} and the impact of the task framing on performance~\citep{kotturi19HirePeer, Levy18contest, chandler13monotony}. The findings we report in this paper can be combined with insights from the aforementioned works to improve the design of the review process with a goal of achieving better efficiency and engagement of reviewers. Additionally, this work is complementary to a direction of research that aims at improving computational and statistical aspects of peer review~\citep{kurokawa15impartial,wang2018your, xu2018strategyproof, Lian18assignment, stelmakh2019testing,fiez2019super,jecmen2020manipulation, noothigattu2018choosing} and results of the present study can be used to motivate future theoretical research.

The rest of the paper is structured as follows. In Section~\ref{section:methodology} we discuss the methodology of each component of the novice-reviewer experiment. We then present the main results in Section~\ref{section:evaluation}. Finally, in Section~\ref{section:discussion} we conclude the paper with a discussion of various aspects of the experiment. 

\section{Methodology}
\label{section:methodology}

In this section we discuss the setup of our experiment. Specifically, we introduce the selection and mentoring mechanisms and explain the methodology of evaluation of reviewers recruited through our experiment in the \ICMLyear{} conference --- a large venue that receives thousands of paper submissions and has more than three thousand reviewers.

\subsection{Selection Mechanism}
\label{section:mech}

The high-level idea of our selection mechanism is to pretest abilities of candidates to write high-quality reviews. To this end, we frame the experiment as an auxiliary peer-review process that mimics the pipeline of the real ML conferences as explained below and ask participants to serve as reviewers in this process. Let us now describe the experiment in detail by discussing the pools of participants and papers, the organization of the auxiliary review process, and the selection criteria we used to identify the best reviews whose authors were invited to join the \ICML{} reviewer pool.

\medskip

\noindent \textbf{Papers.} We solicited $19$ anonymized preprints in various sub-areas of ML from colleagues at various research labs, ensuring that authors of these manuscripts do not participate in the experiment as subjects. Some ML and AI conferences publicly release reviews for accepted/submitted papers, making these papers inappropriate for our experiment as our goal is to elicit independent reviews from participants. Thus, we used only those papers that did not have reviews publicly available. The final pool of papers consisted of working papers, papers under review at other conferences, workshop publications and unpublished manuscripts. The papers were 6--12 pages long excluding references and appendices (a standard range for many ML conferences) and were formatted in various popular journals' and conferences' templates with all explicit venue identifiers removed.

\medskip

\noindent \textbf{Participants.} Since we had a small quota of approximately 50 reviewers allocated for the experiment in the reviewer pool of the \ICMLyear{} conference, in this positional experiment we limited the target study population to graduate students or recent graduates of five large, top US universities (CMU, MIT, UMD, UC Berkeley and Stanford). To recruit participants, we messaged mailing lists of these universities and targeted master's and junior PhD students working in ML-related fields. The invitation also propagated to a small number of students outside of these schools through the word of mouth. The recruiting materials contained an invitation to participate in the \ICML{} reviewer-selection experiment. Specifically, we notified participants that they will need to review one paper and that those who write strong reviews will be invited to join the the \ICML{} reviewer pool. Being a reviewer in the top ML conference is a recognition of one's expertise and we envisaged that this potential benefit is a good motivation for junior researchers to join our experiment. As a result, we received responses from $200$ candidates, more than 90\% of whom were students/recent graduates from the aforementioned schools. All of these candidates were added to the pool of participants without further screening. We provide additional discussion of the demography of participants (including their research and reviewing experience) in Section~\ref{section:discussion}.

\medskip

\noindent \textbf{Auxiliary peer-review process} The selection procedure closely followed the initial stages of the standard double-blind ML conference peer-review pipeline and was hosted using Microsoft Conference Management Toolkit (\url{https://cmt3.research.microsoft.com}) which is also used in \ICML{}.
First, we asked participants to indicate their preferences in what papers they would like to review by entering bids that take the following values: ``Not Willing'', ``In a Pinch'', ``Willing''  and ``Eager''. Thirteen participants did not enter any bids and were removed from the pool. The remaining 187 participants were active in bidding (mean number of ``Willing''  and ``Eager'' bids is 4.7) and we assigned all of them to 1 paper each, where we tried to satisfy reviewer bids, subject to a constraint that each paper is assigned to at least 8 reviewers.\footnote{The constraint on the number of reviewers per paper was enforced to facilitate another experiment conducted in parallel with the present study and described in the companion paper~\citep{stelmakh2020outcome}.}  As a result, 186 participants were assigned to a paper they bid either ``Willing'' or ``Eager'' and 1 participant was assigned to a paper they bid ``In a Pinch'' (this participant did not bid ``Eager'' or ``Willing'' on any paper).

Finally, we instructed participants that they should review the paper as if it was submitted to the real \ICML{} conference with the exception that the relevance to \ICML{}, formatting issues (e.g., page limit, margins) and potential anonymity issues should not be considered as criteria. To help participants in writing their reviews, we provided reviewer guidelines (included in supplementary materials on the first author's website) that discuss the best practices of reviewing. We gave participants 15 days to complete the review and then extended the deadline for 16 more days to accommodate late reviews as our original deadline interfered with the final exams at various US universities and the US holiday period.

\smallskip

\noindent \textbf{Selection of participants} Out of 187 participants who were assigned a paper for review, 134 handed in the reviews (response rate of $71.7\%$). Upon receipt of reviews, we removed numeric scores given by participants to the papers and relied on the combination of the following approaches to identify individuals to be invited to join the \ICML{} reviewer pool:

\begin{itemize}[itemsep=1pt, leftmargin=15pt, topsep=2pt]
    \item \textbf{Author evaluation} We asked authors of papers used in the experiment to read the reviews and rate/comment on their qualities. Authors of 14 of the 19 submissions responded to our request.    
    
    \item \textbf{Internal evaluation} We analyzed reviews for 17 papers falling in the study team members' areas of expertise.
    
    \item \textbf{External evaluation} We called upon an independent domain expert to help with 2 papers that are outside of the study team members' areas of expertise.
\end{itemize}

It is natural to assume that authors are at the best position to evaluate the reviews written for their papers. Indeed, they know all the technical details of their papers, thereby being able to evaluate objective points made by reviewers. Additionally, we hypothesize that the non-competitive nature of the auxiliary review process may reduce potential biases related to a more negative perception of critical reviews which in the past were observed in some real conferences~\citep{weber2002author, apagiannaki2007pam, khosla2012cvpr}. With this motivation, we requested authors to provide feedback on the received reviews and most of the authors fulfilled our request.

To validate our expectations regarding the quality of the author feedback, all the reviews together with the author feedback (when available) were additionally analyzed by the study team members and the aforementioned domain expert. We qualitatively observed that the author feedback is helpful to identify the strongest reviews and our selection decisions were well-aligned with the authors' evaluations. Overall, we invited 52 participants whose reviews received excellent feedback from all the evaluators who read the review to join the \ICML{} reviewer pool; all 52 accepted the invitation. For the rest of the paper, we will refer to these reviewers as \exprev{} reviewers.

\subsection{Mentoring Mechanism} 
\label{section:mentoring}

Throughout the conference review process, the \exprev{} reviewers were offered additional mentorship:
\begin{itemize}[itemsep=1pt, leftmargin=15pt, topsep=2pt]
    \item The reviewers were provided with a senior researcher as a point of contact, and were offered to ask any questions pertaining to the review process at any point in the process. There were several questions asked and answered as a part of the mentorship.
    
    \item The reviewers were provided with examples on various parts of the process, for instance, on how to  lead a discussion among the reviewers.

    \item When the initial reviews were submitted, certain issues were identified that were common across many reviews from the \exprev{} pool (e.g., many reviews were initially written about the authors rather than the paper). The \exprev{} reviewers were requested to address these issues.
    
    \item The \exprev{} reviewers were sent a few more reminders than the conventional reviewers. 
\end{itemize}
The total amount of time and effort in the mentorship (across all 52 \exprev{} reviewers) was equal to about half the time and effort for a meta-reviewer's job.

\subsection{Methodology of Evaluation}
\label{section:meva}

The main pool of the \ICML{} 2020 reviewers was recruited through a combination of conventional approaches and consisted of 3,012 reviewers\footnote{Some reviewers who initially accepted the invitation dropped out in the early stages of the review process and are not included in this number and in the subsequent analysis.} belonging to two disjoint groups. The first group, which we refer to as \currev{}, made up about 68\% of the main pool and included reviewers who were invited by program chairs based on satisfaction of at least one of the following criteria: (i) several years of reviewing and publishing experience for top ML venues, (ii) above-average performance in reviewing for NeurIPS 2019 or (iii) personal recommendation by a meta-reviewer. The remaining 32\% of reviewers constituted the second group that we call \snrev{}: this group comprised individuals who self-nominated and satisfied the selection criteria of (i) having at least two papers published in some top ML venues, and (ii) being a reviewer for at least one top ML conference in the past.  On average, the \currev{} group consisted of more senior researchers while the \snrev{} pool mostly comprised researchers at early stages of their careers.

In the sequel, we compare the performance of 52 \exprev{} reviewers who joined the \ICML{} reviewer pool through our experiment with the performance of the reviewers from the main pool.  Let us now discuss some important details of the evaluation.

\medskip

\noindent \textbf{Affiliation caveat} 51 out of 52 \exprev{} reviewers recruited through our selection procedure are current master's and PhD students or recent graduates of the aforementioned universities (one reviewer is a graduate of another US school), whereas reviewers from the main pool represent universities as well as private companies, government organizations, non-profits and more, from all over the world. Hence, the reviewers in the main pool have different backgrounds from the \exprev{} reviewers and this difference can serve as an undesirable confounder (orthogonal to the selection procedure and mentoring) in our analysis.

To counteract this confounding factor, we identify  a subset of the main pool of reviewers, whom we call \elrev{} reviewers. The \elrev{} group comprises 305 reviewers from the main pool who share an affiliation (i.e., email domain or affiliation listed on the conference management system) with the 5 schools mentioned above. In our evaluations subsequently, we additionally juxtapose the \exprev{} reviewers to this group to evaluate how they compare to reviewers of similar background, thereby alleviating the affiliation confounder.

\medskip

\noindent \textbf{Metrics and tools of comparison} We use a set of indirect indicators of review quality (e.g., review length and discussion participation) as well as direct evaluations of review quality made by meta-reviewers of the \ICML{} conference ---  senior reviewers, each of whom is in charge of overseeing the review process for approximately 20 submissions. To quantify significance of the difference in these metrics, we use the permutation test~\citep{fisher35permutation}, treating each paper-reviewer pair as a unit of analysis. Error bars presented in figures below represent bootstrapped 90\% confidence intervals unless stated otherwise.

Finally, throughout the review process, meta-reviewers were calling upon additional external reviewers to help with some submissions or asking reviewers from the main pool to review additional papers; these paper-reviewer pairs are not included into comparison because new reviewers typically had less time to complete the assigned reviews.

\section{Evaluation}
\label{section:evaluation}

In the previous section we described our approach towards recruiting novice reviewers and mentoring them. In this section we move to the real \ICML{} conference and evaluate the benefit of the proposal by juxtaposing the performance of \exprev{} reviewers to the main reviewer pool which consists of \snrev{} and \currev{} reviewers, some of whom belong to the group of \elrev{} reviewers. For this, we compare performance of reviewers at different stages of the review process: bidding, reviewing (in-time submission, review length, self-assessed confidence and others) and discussion (activity, attention to the author feedback). Finally, we complement the comparison by overall evaluation of the review quality made by meta-reviewers.

\begin{table*}[t]
\begin{center}
\begin{footnotesize}
\begin{sc}
\begin{tabular}{lcccr}
\toprule
   Criteria (R = reviewer, P = paper)     & Range & Experimental & Main Pool & $P$ val. \\
\midrule
1.* Mean number of positive bids per R  & $[0, 5052]$ & ${34.6}$ & $27.4$ & $.043$ \\
2.* Frac. of Rs with $>$ 0 reviews completed in time & [0, 1] & ${0.92}$ & $0.81$ & $.041$ \\
3.* Mean review length (in symbols) & $[0, \infty)$ & ${4759}$ & $2858$ & $<.001$ \\
4.\phantom{*} Mean initial overall score given by Rs & $[1, 6]$ & $3.34$ & $3.25$ & $.373$ \\
5.\phantom{*} Mean self-reported confidence & $[1, 4]$ & $3.05$ & $3.03$ & $.841$ \\
6.* Mean self-reported expertise & $[1, 4]$ & $2.83$ & $2.98$ & $.026$ \\
7.* Frac. of (P, \!R) pairs with R active in P discussion & $[0, 1]$ & $0.68$ & $0.58$ & $.033$ \\
8.* Frac. of (P, \!R) pairs with post-rebuttal review update & $[0, 1]$ & $0.61$ & $0.43$ & $<.001$ \\
9.* Mean review quality evaluated by meta-R & $[1, 3]$ & ${2.26}$ & $2.08$ & $<.001$ \\
\bottomrule
\end{tabular}
\end{sc}
\end{footnotesize}
\end{center}
\caption{Performance comparison of reviewers from the main pool and \exprev{} reviewers on various criteria. Asterisks indicate criteria with significant difference at the level 0.05.} 
\label{table:joint}
\end{table*}

Table~\ref{table:joint} summarizes the results of comparison of \exprev{} reviewers with reviewers from the main pool; subsequently, we will present a more detailed analysis with breakdown by reviewer groups. The main message of Table~\ref{table:joint} is that from various angles the reviews written by \exprev{} reviewers are comparable to or sometimes even better than reviews written by reviewers from the main pool. With this general observation, we now provide details and background for each row of Table~\ref{table:joint}.

   
\paragraph{Bidding activity (Row 1 of Table~\ref{table:joint})} Algorithms for automated paper-reviewer matching significantly rely on reviewer bids~\citep{fiez2019super}. Hence, activity of reviewers in the bidding stage is crucial to ensure that submissions are assigned to reviewers with appropriate expertise. To give matching algorithms enough flexibility, \ICML{} program chairs requested reviewers to positively bid (i.e., indicate papers they are ``Willing'' or ``Eager'' to review) on at least 30-40 submissions (out of approximately 5,000 submitted for review).

\begin{figure}[t]
    \centering
    \includegraphics[width=8cm]{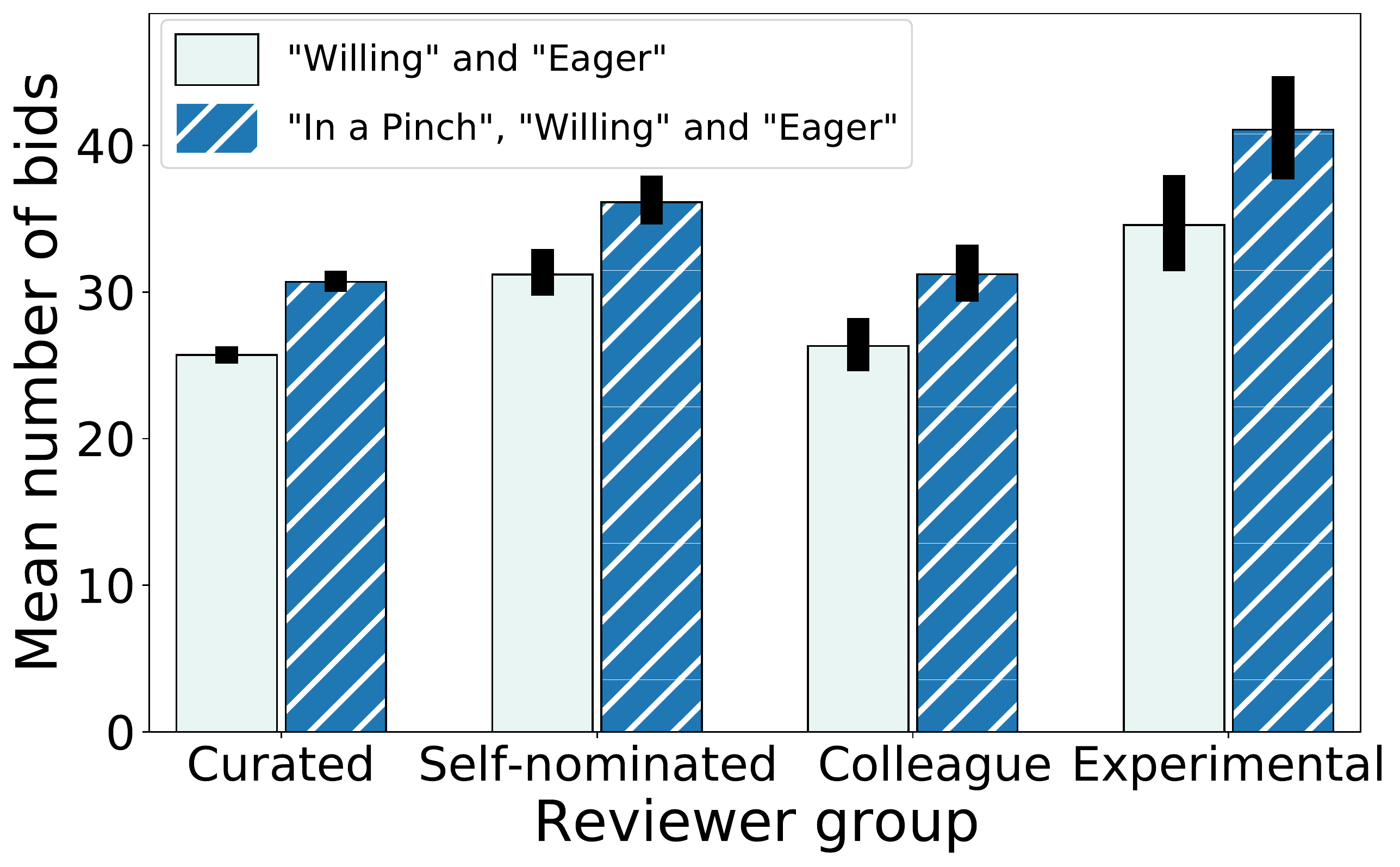}
    \caption{Mean number of positive/non-negative bids per reviewer. {\sc Experimental} reviewers positively bid on more papers than reviewers from each of the comparison groups.}
    \label{fig:bidding}
\end{figure}

Figure~\ref{fig:bidding} compares mean numbers of positive and non-negative (``Willing'', ``Eager''  and ``In a Pinch'') bids made by reviewers from different groups. Note that to compare non-negative bids we remove one reviewer who bulk bid ``In a Pinch'' on all non-conflicting submissions. Several reviewers bid on hundreds of submissions (possibly by bulk bidding on some specific areas or keywords), but we do not exclude such reviewers from the analysis.

Overall, we observe that \exprev{} reviewers are more active than other categories of reviewers with a qualification that the difference with \snrev{} reviewers ($\Delta_{\text{positive}} = 3.4, \ \Delta_{\text{non-negative}} = 5.0$) is not statistically significant at the 0.05 significance level as demonstrated in Table~\ref{table:bidding} that summarizes the results of comparison.\footnote{We also observe that \snrev{} reviewers are more active than \currev{} reviewers, suggesting that the bidding activity may be decreasing with seniority.} 

\begin{table*}[t]
\begin{center}
\begin{small}
\begin{sc}
\begin{tabular}{llccccc}
\toprule
  Bids &  & Experimental & Main Pool & Curated & Self-Nominated & Colleague \\
\midrule
\multirow{4}{*}{Positive} & Sample Size & $52$ & $3012$ & $2060$ & $952$ & $305$ \\
                               & Mean Value & $34.6$ & $27.4$ & $25.7$ & $31.2$ & $26.3$ \\
                               & 90\% CI & $[31.4; 38.1]$ & $[26.8; 28.1]$ & $[25.1; 26.3]$ & $[29.8; 33.0]$ & $[24.6; 28.3]$ \\
                               & $P$ value  & -- & $.043$ & $.003$ & $.221$ & $.011$ \\
\midrule
 & Sample Size & $52$ & $3011$ & $2059$ & $952$ & $305$ \\
 {Non-} & Mean Value  & $41.1$ & $32.4$ & $30.7$ & $36.1$ & $31.2$ \\
 {Negative} & 90\% CI &  $[37.6; 44.7]$ & $[31.7; 33.2]$ & $[30.0; 31.5]$ & $[34.6; 37.9]$ & $[29.3; 33.3]$\\
                               & $P$ value  & -- & $.046$ & $.007$ & $.165$ & $.005$\\
\bottomrule
\end{tabular}
\end{sc}
\end{small}
\end{center}
\caption{Comparison of bidding activity of the reviewers. $P$ values are for the test of the difference of means between \exprev{} and each of the other groups of reviewers.} 
\label{table:bidding}
\end{table*}

\begin{figure}[t]
\centering
\begin{subfigure}[t]{0.48\textwidth}%
    \vskip 0pt
    \centering
    \includegraphics[width=7cm]{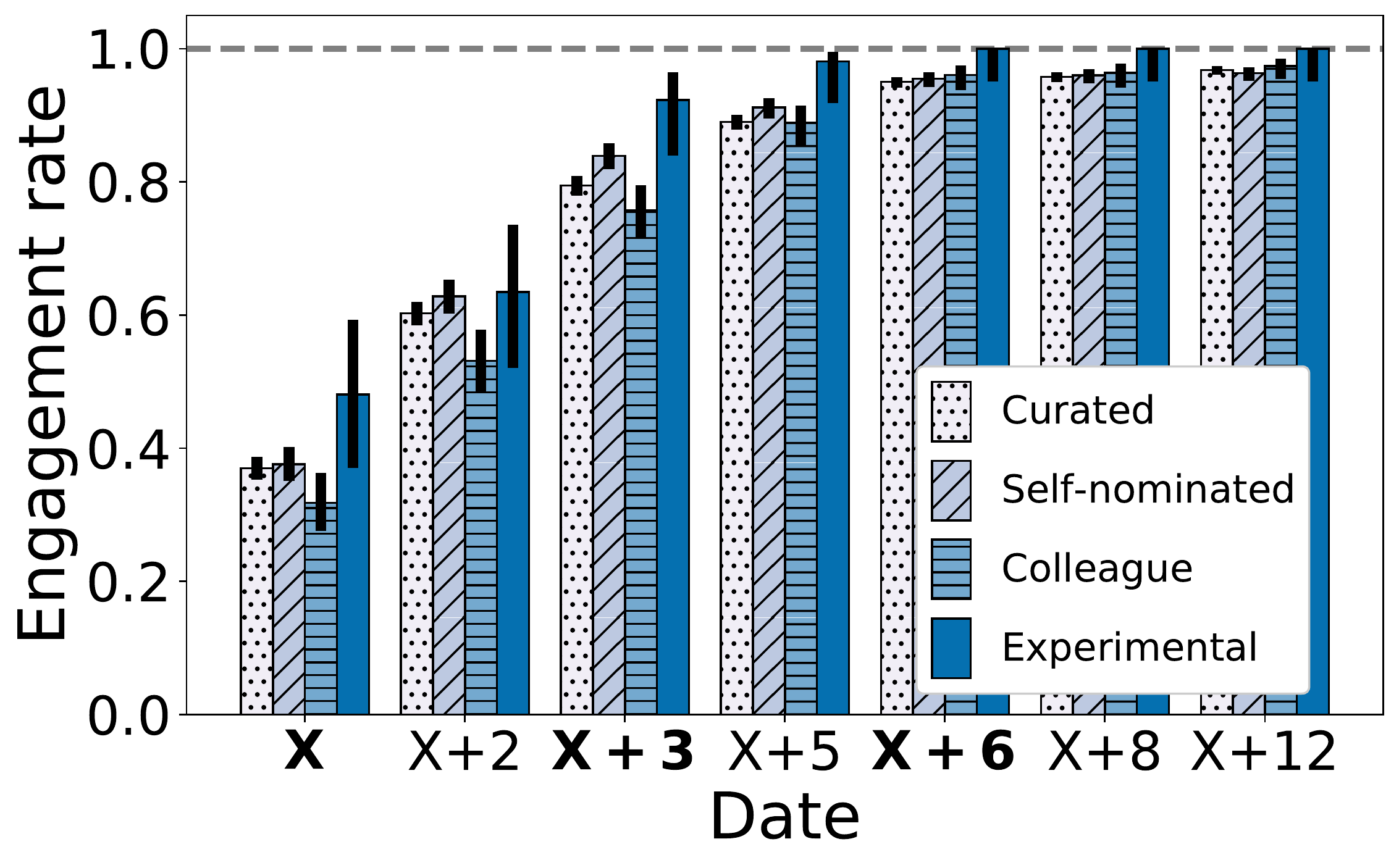}
    \caption{Engagement rates.}
    \label{fig:completion}
\end{subfigure}\hspace{5pt}%
\begin{subfigure}[t]{0.48\textwidth}%
    \vskip 0pt
    \centering
    \includegraphics[width=7cm]{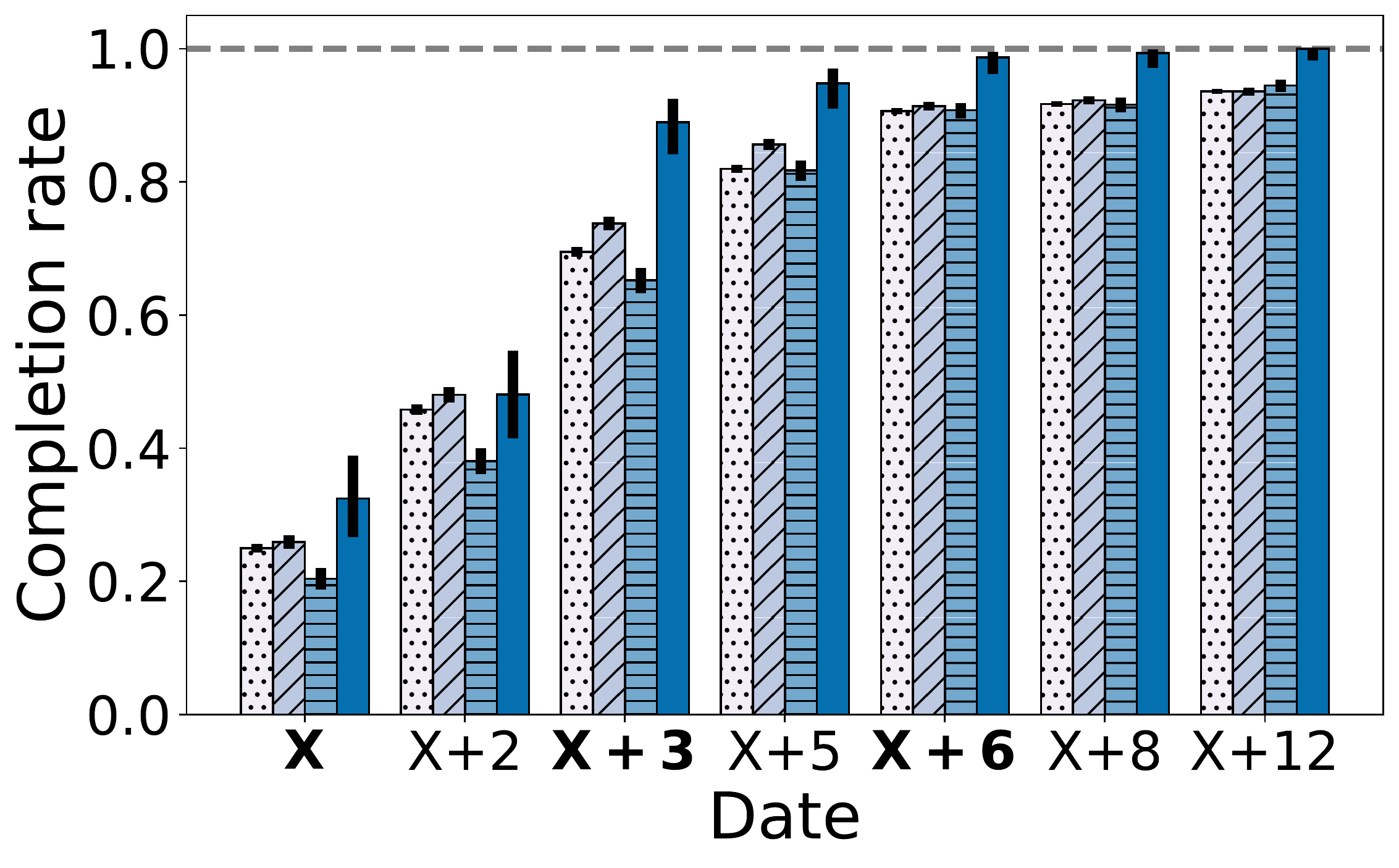}
        \caption{Completion rates.}
    \label{fig:completion_rate}
\end{subfigure}
\caption{Timely review submission. Bold labels indicate dates at which deadlines were set with the original deadline on day X and two extensions. 90\% confidence intervals are computed using the method of~\citet{wilson27test}. {\sc Experimental} reviewers have higher engagement and completion rates than other reviewers.} 
\label{fig:intime}
\end{figure}

\paragraph{Timely review submission (Row 2 of Table~\ref{table:joint})} A typical conference timeline is very tight and it is crucial that reviewers complete their reviews in a timely manner. We now compare how different groups of reviewers respect the deadlines. For this, we use two metrics: first, Figure~\ref{fig:completion} juxtaposes engagement rates --- fractions of reviewers who submitted at least one review by a given date --- of different reviewer groups. Second, Figure~\ref{fig:completion_rate} compares completion rates --- the total number of submitted reviews divided by the total number of assigned papers. While the completion rate is perhaps a more intuitive choice of metric, it is artificially favourable to \exprev{} reviewers due to  a difference in the reviewer loads between \exprev{} reviewers and reviewers from the main pool (see more discussion in Section~\ref{section:discussion}). To counteract the bias in objective, for formal comparisons presented in Tables~\ref{table:joint} and~\ref{table:timing} we use the engagement rate which is less impacted by the difference in loads.

Looking at Figure~\ref{fig:intime}, we again observe the trend of junior \snrev{} and \exprev{} reviewers being consistently more active than their senior \currev{} counterparts throughout the whole review-submission period, with \exprev{} reviewers achieving the highest engagement and completion rates across all reviewer groups. Note that due to the impact of the COVID-19 pandemic, the initial deadline for review submission on day X was extended twice (deadline dates are highlighted in bold in Figure~\ref{fig:intime}) with the first extension announced well in advance and hence only a small fraction of reviews was submitted by day X. Thus, in Table~\ref{table:joint} we use data for the first extended deadline on day X+3. 

\begin{table*}[t]
\begin{center}
\begin{small}
\begin{sc}
\begin{tabular}{lccccr}
\toprule
 & Experimental & Main Pool & Curated & Self-Nominated & Colleague \\
\midrule
Sample Size & $52$ & $3007$ & $2055$ & $952$ & $305$ \\
Mean Value & $0.92$ & $0.81$ & $0.79$ & $0.84$ & $0.76$ \\
90\% Wilson CI & $[0.84; 0.96]$ & $[0.80; 0.82]$ & $[0.78; 0.81]$ & $[0.82; 0.86]$ & $[0.71; 0.80]$ \\
$P$ value  & -- & $.041$ & $.028$ & $.140$ & $.006$ \\
\bottomrule
\end{tabular}
\end{sc}
\end{small}
\end{center}
\caption{Comparison of engagement rates of reviewers on the first extended deadline. $P$ values are for the test of the difference of means between \exprev{} and each of the other groups of reviewers.} 
\label{table:timing}
\end{table*}

Table~\ref{table:timing} extends the comparison reported in Table~\ref{table:joint} by displaying a breakdown by reviewer groups.\footnote{The number of reviewers used in the comparison is smaller than the total number of reviewers because we only use paper-reviewer pairs that were in the assignment from the beginning of the review period and 5 reviewers from the \currev{} group with small initial loads had a set of their papers fully changed throughout the process.} The results of the permutation test qualify the observations we made from Figure~\ref{fig:intime} by showing that the difference between \exprev{} and \snrev{} reviewers ($\Delta = 0.08$), who represent the more junior population of the main reviewer pool, is not significant at the requested level.

Before we proceed to other dimensions of comparison, we note that a small number of reviewers from the main pool never submitted reviews for some of the assigned papers (less than 5\% of paper-reviewer pairs had no review submitted). Corresponding paper-reviewer pairs are excluded from the analysis of various aspects of review quality we perform below.

\begin{figure}[t]
    \centering
    \includegraphics[width=7cm]{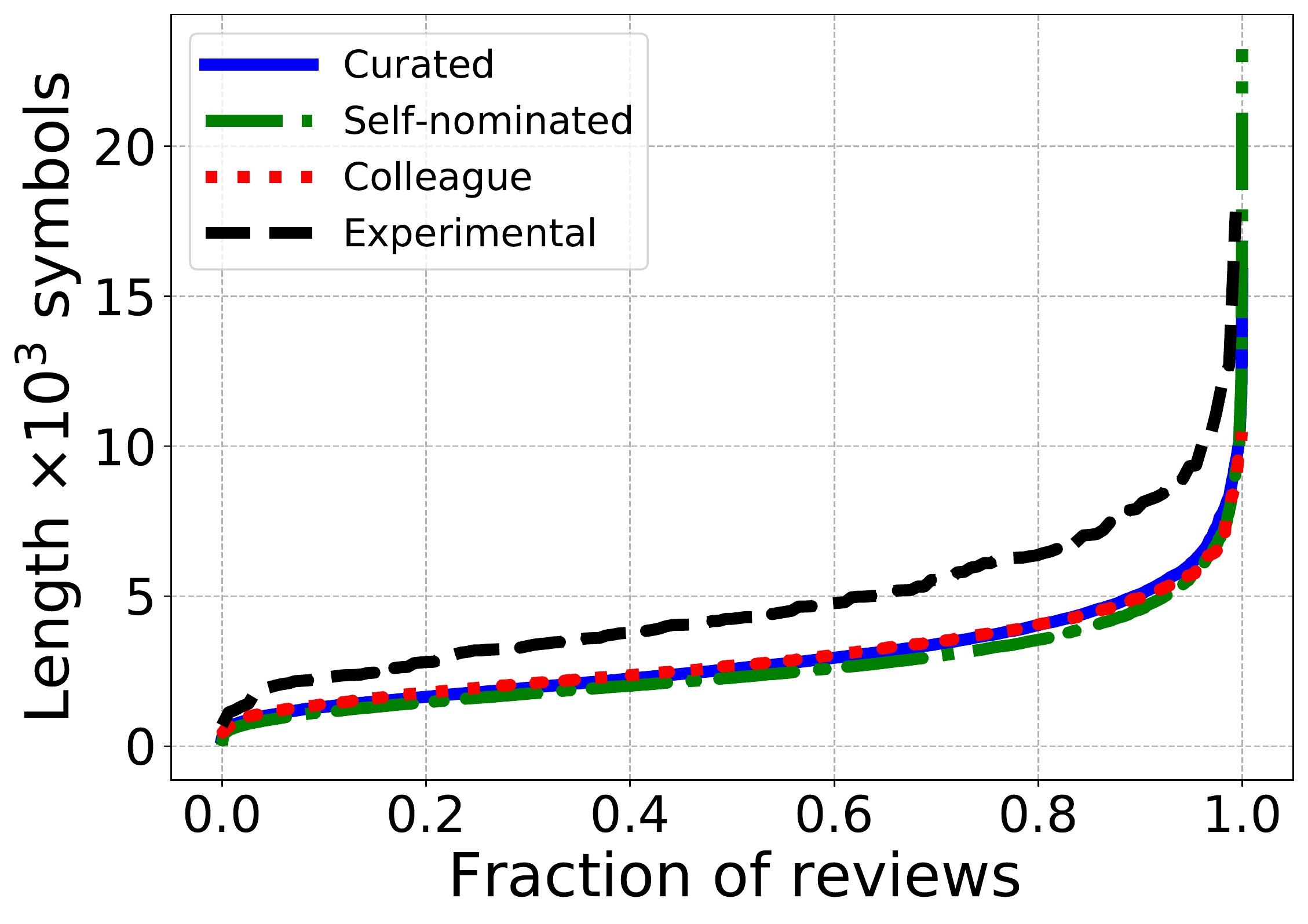}
    \caption{Distribution of review lengths. {\sc Experimental} reviewers write longer reviews than other reviewers.}
    \label{fig:revlen}
\end{figure}

\paragraph{Review length (Row 3 of Table~\ref{table:joint})} We continue the analysis by juxtaposing the lengths of textual comments submitted by reviewers in Figure~\ref{fig:revlen}. We observe that different categories of reviewers from the main pool appear to write reviews of comparable length whereas \exprev{} reviewers write considerably longer reviews. The distribution of lengths of reviews written by reviewers from the main pool is very similar to that of several major ML conferences~\citep{beygelzimer19data}, and thus we conclude that \exprev{} reviewers produced longer reviews than standard in the field. Table~\ref{table:len} compares mean lengths of reviews written by reviewers from different groups and confirms the intuition represented in Figure~\ref{fig:revlen}. 

\begin{table*}[t]
\begin{center}
\begin{small}
\begin{sc}
\begin{tabular}{lccccr}
\toprule
   & Experimental & Main Pool & Curated & Self-Nominated & Colleague \\
 \midrule
 Sample Size & $154$ & $15206$ & $10502$ & $4704$ & $1593$ \\
                               Mean Value & $4759$ & $2858$ & $2953$ & $2647$ & $2985$ \\
                               90\% CI & $[4432; 5089]$ & $[2836; 2880]$ & $[2926; 2979]$ & $[2609; 2685]$ & $[2924; 3047]$ \\
                               $P$ value  & -- & $<.001$ & $<.001$ & $<.001$ & $<.001$ \\
\bottomrule
\end{tabular}
\end{sc}
\end{small}
\end{center}
\caption{Comparison of mean lengths (in symbols) of reviews. $P$ values are for the test of the difference of means between \exprev{} and each of the other groups of reviewers.} 
\label{table:len}
\end{table*}


\paragraph{Hypercriticality (Row 4 of Table~\ref{table:joint})}

\begin{figure}[b]
\centering
\includegraphics[width=8cm]{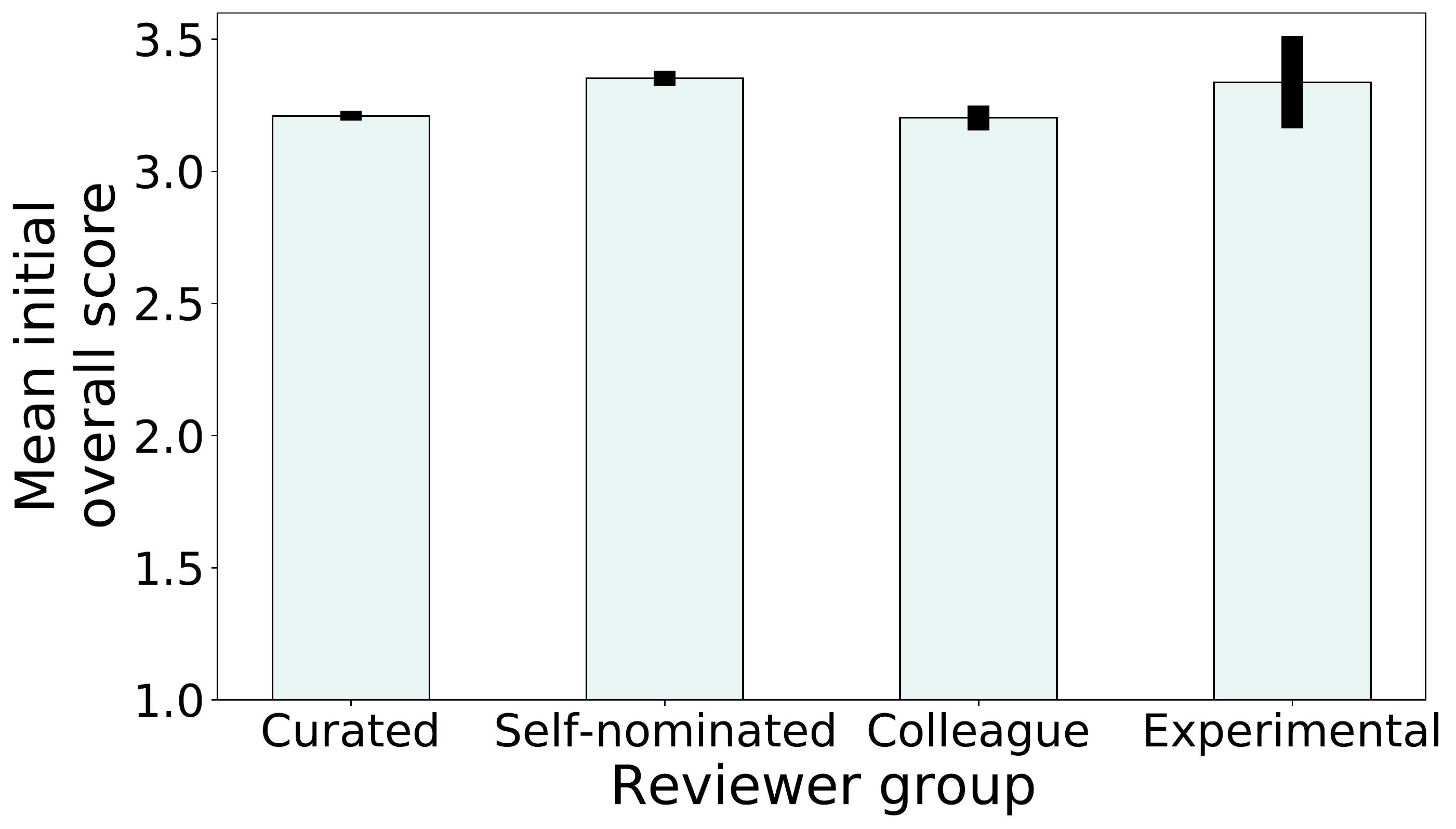}
\caption{Mean initial overall scores. {\sc Self-nominated} and \exprev{} reviewers appear to be more lenient than \currev{} reviewers.}
\label{fig:mean}
\end{figure}

Although junior reviewers are often perceived to be more critical~\citep{mogul13hypercriticality, tomiyama07hypercticality, toor09hypercticality}, the analysis of the NeurIPS 2016 conference  conducted by~\citet{shah2017design} does not reveal a significant difference between overall scores given by junior and senior reviewers. We now perform such an analysis for the \ICML{} conference: in \ICMLyear{} reviewers were asked to give the overall score on a 6-point Likert item and we encode the options with integers from 1 to 6 such that the larger number indicates the better score.  The mean overall scores given in \emph{initial} reviews (i.e., before reviewers got to see other reviews and the author rebuttal) are compared across different groups of reviewers in Figure~\ref{fig:mean}. 

Mean initial overall scores given by different groups of reviewers appear to be comparable; \snrev{} and \exprev{} reviewers seem to be slightly more lenient than \currev{} reviewers, but the sample size of \exprev{} reviewers is not sufficient to draw definitive conclusions (Table~\ref{table:hyper} summarizes the comparison). However, we note that \snrev{} reviewers are indeed more lenient than \currev{} reviewers  ($\Delta = 0.14, P < .001$), contradicting the aforementioned observations of hypercriticality in junior reviewers. This misalignment may be specific to the field of computer science where hypercriticality is not limited to junior reviewers, but is prevalent in the whole area~\citep{vardi2010hypercriticality}, or, alternatively, it is possible that hypercriticality of junior reviewers manifests not in numeric scores but in textual reviews. 

\begin{table*}[t]
\begin{center}
\begin{small}
\begin{sc}
\begin{tabular}{lccccr}
\toprule
  & Experimental & Main Pool & Curated & Self-Nominated & Colleague \\
\midrule
 Sample Size & $154$ & $15206$ & $10502$ & $4704$ & $1593$ \\
                               Mean Value & $3.34$ & $3.25$ & $3.21$ & $3.35$ & $3.20$ \\
                               90\% CI & $[3.17; 3.51]$ & $[3.24; 3.27]$ & $[3.19; 3.23]$ & $[3.33; 3.38]$ & $[3.16; 3.25]$ \\
                               $P$ value  & -- & $.373$ & $.199$ & $.895$ & $.175$ \\
\bottomrule
\end{tabular}
\end{sc}
\end{small}
\end{center}
\caption{Comparison of mean initial overall scores. $P$ values are for the test of the difference of means between \exprev{} and each of the other groups of reviewers.} 
\label{table:hyper}
\end{table*}


\paragraph{Expertise and confidence (Rows 5 and 6 of Table~\ref{table:joint})}

\begin{figure}[b]
\centering
\begin{subfigure}[t]{0.48\textwidth}%
    \vskip 0pt
    \centering
    \includegraphics[width=7.5cm]{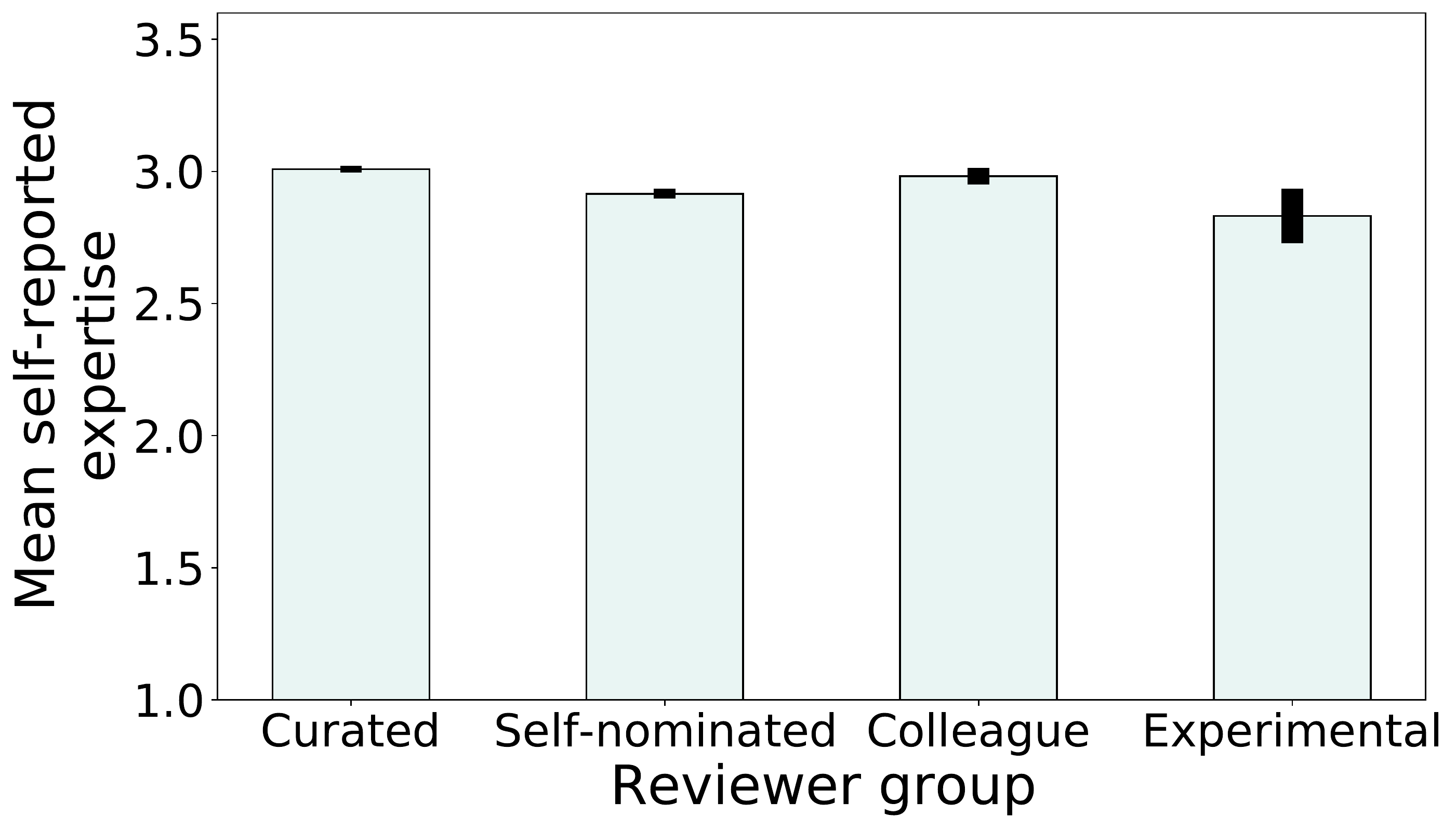}
    \caption{Self-assessed expertise.}
    \label{fig:exp_conf:exp}
\end{subfigure}\hspace{5pt}%
\begin{subfigure}[t]{0.48\textwidth}%
    \vskip 0pt
    \centering
    \includegraphics[width=7.5cm]{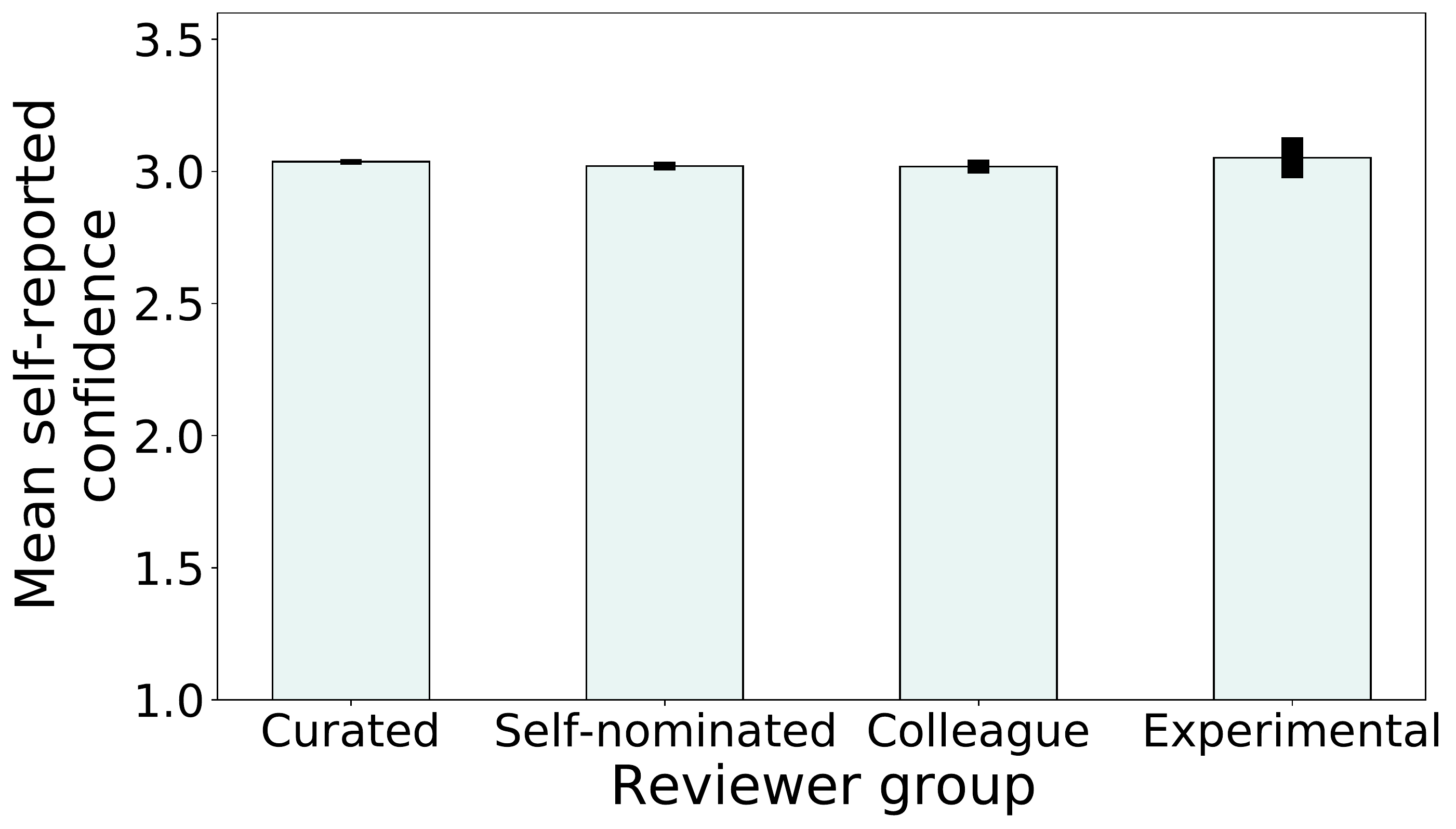}
    \caption{Self-assessed confidence.}
    \label{fig:exp_conf:conf}
\end{subfigure}
\caption{Comparison of self-assessed expertise and confidence. {\sc Experimental} reviewers report considerably lower expertise than other groups of reviewers, but there is no significant difference in self-assessed confidence.} 
\label{fig:exp_conf}
\end{figure}

We now continue with the analysis of self-assessed confidence and expertise of different reviewer groups (using values reported in initial reviews). The reviewer form of the \ICMLyear{} conference contained two questions in which reviewers were asked to evaluate their expertise and confidence in their review on 4-point Likert items. We encode the options of the Likert items with integer numbers from 1 to 4 such that larger numbers indicate higher expertise/confidence. We then compare mean scores across reviewer groups and report the results in Figure~\ref{fig:exp_conf} and Table~\ref{table:conf_and_exp}.

Not surprisingly, \exprev{} reviewers reported lower self-assessed expertise (Figure~\ref{fig:exp_conf:exp}) with a caveat that the difference with \snrev{} reviewers is not statistically significant. That said, the difference in expertise does not seem to result in the difference in self-assessed confidence (Figure~\ref{fig:exp_conf:conf}). 

\begin{table*}[t]
\begin{center}
\begin{small}
\begin{sc}
\begin{tabular}{llccccr}
\toprule
  Criteria &  & Experimental & Main Pool & Curated & Self-Nominated & Colleague \\
\midrule
\multirow{4}{*}{Expertise} & Sample Size & $154$ & $15206$ & $10502$ & 4704 & 1593 \\
                               & Mean Value & $2.83$ & $2.98$ & $3.01$ & $2.92$ & $2.98$ \\
     & 90\% CI & $[2.73; 2.94]$ & $[2.97; 2.99]$ & $[3.00; 3.02]$ & $[2.90; 2.93]$ & $[2.95; 3.01]$ \\
                               & $P$ value  & -- & $.026$ & $.005$ & $.204$ & $.021$ \\
\midrule
\multirow{4}{*}{Confidence} & Sample Size & $154$ & $15206$ & $10502$ & 4704 & 1593 \\
                               & Mean Value  & $3.05$ & $3.03$ & $3.04$ & $3.02$ & $3.02$ \\
                               & 90\% CI  & $[2.97; 3.13]$ & $[3.02; 3.04]$ & $[3.03; 3.05]$ & $[3.00; 3.04]$ & $[2.99; 3.05]$ \\
                               & $P$ value  & -- & $.841$ & $.829$ & $.646$ & $.579$ \\
\bottomrule
\end{tabular}
\end{sc}
\end{small}
\end{center}
\caption{Comparison of self-assessed expertise (first 4 rows)  and confidence (last 4 rows). $P$ values are for the test of the difference of means between \exprev{} and each of the other groups of reviewers.} 
\label{table:conf_and_exp}
\end{table*}


\paragraph{Rebuttals and discussion (Rows 7 and 8 of Table~\ref{table:joint})} The review process of \ICML{} allows authors to respond to initial reviews written for their papers by submitting a short rebuttal that is followed by a private discussion between reviewers and the meta-reviewer. Past analysis~\citep{shah2017design,gao19rebuttal,aclpc18data} provide mixed evidence regarding the usefulness of rebuttals, and in this work we do not aim to judge the overall efficacy of the rebuttal process. However, in order for the rebuttal or discussion to change the reviewer's opinion, reviewers at the very least need to consider the rebuttal and be engaged in the discussion. We now investigate this aspect, conditioning on papers whose authors supplied a response to initial reviews (approximately 80\% of submissions had the author response provided).

\begin{figure}[b]
\centering
\begin{subfigure}[t]{0.48\textwidth}%
    \vskip 0pt
    \centering
    \includegraphics[width=7.5cm]{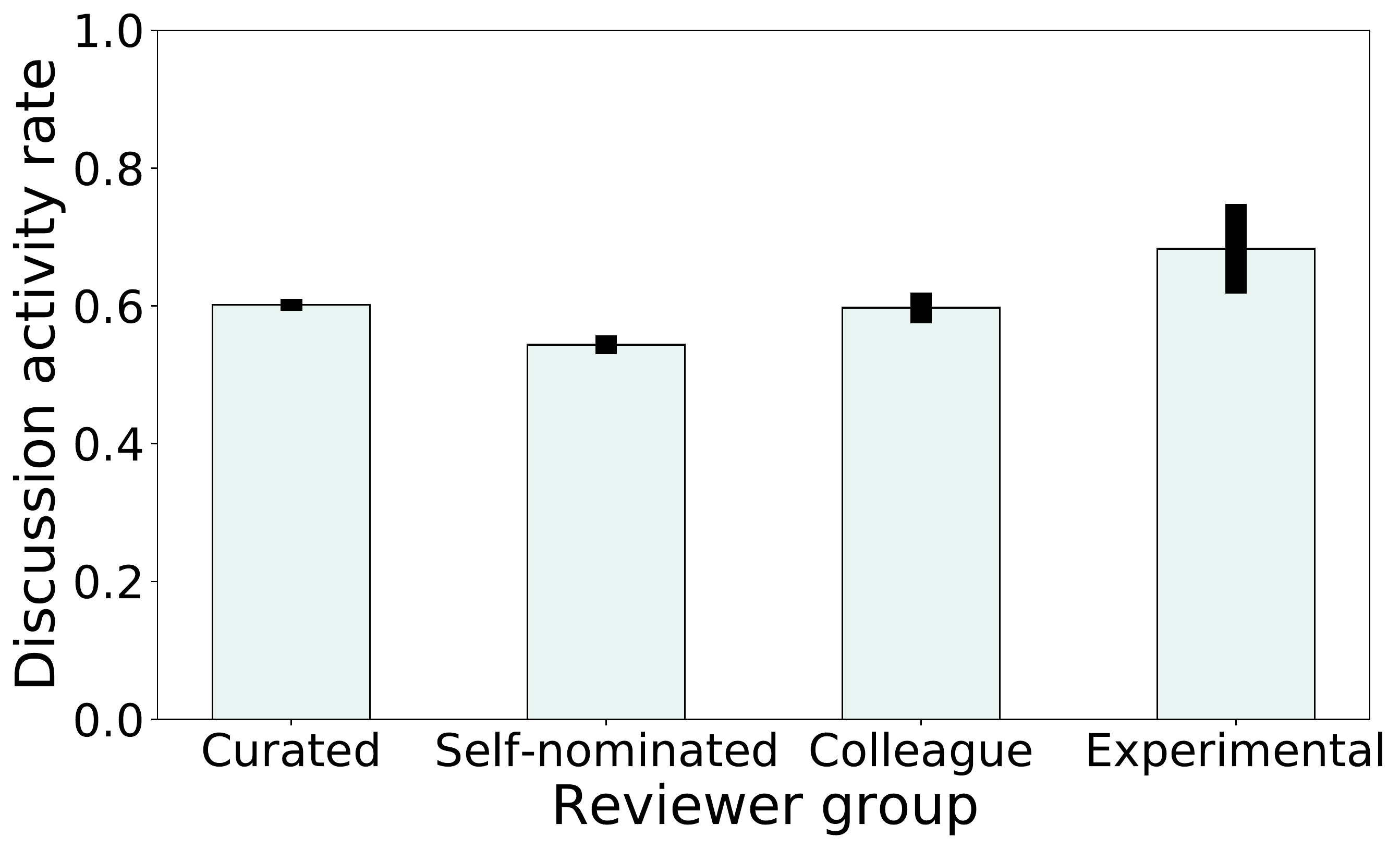}
    \caption{Participation in discussion.}
    \label{fig:discussion:disc}
\end{subfigure}\hspace{5pt}%
\begin{subfigure}[t]{0.48\textwidth}%
    \vskip 0pt
    \centering
    \includegraphics[width=7.5cm]{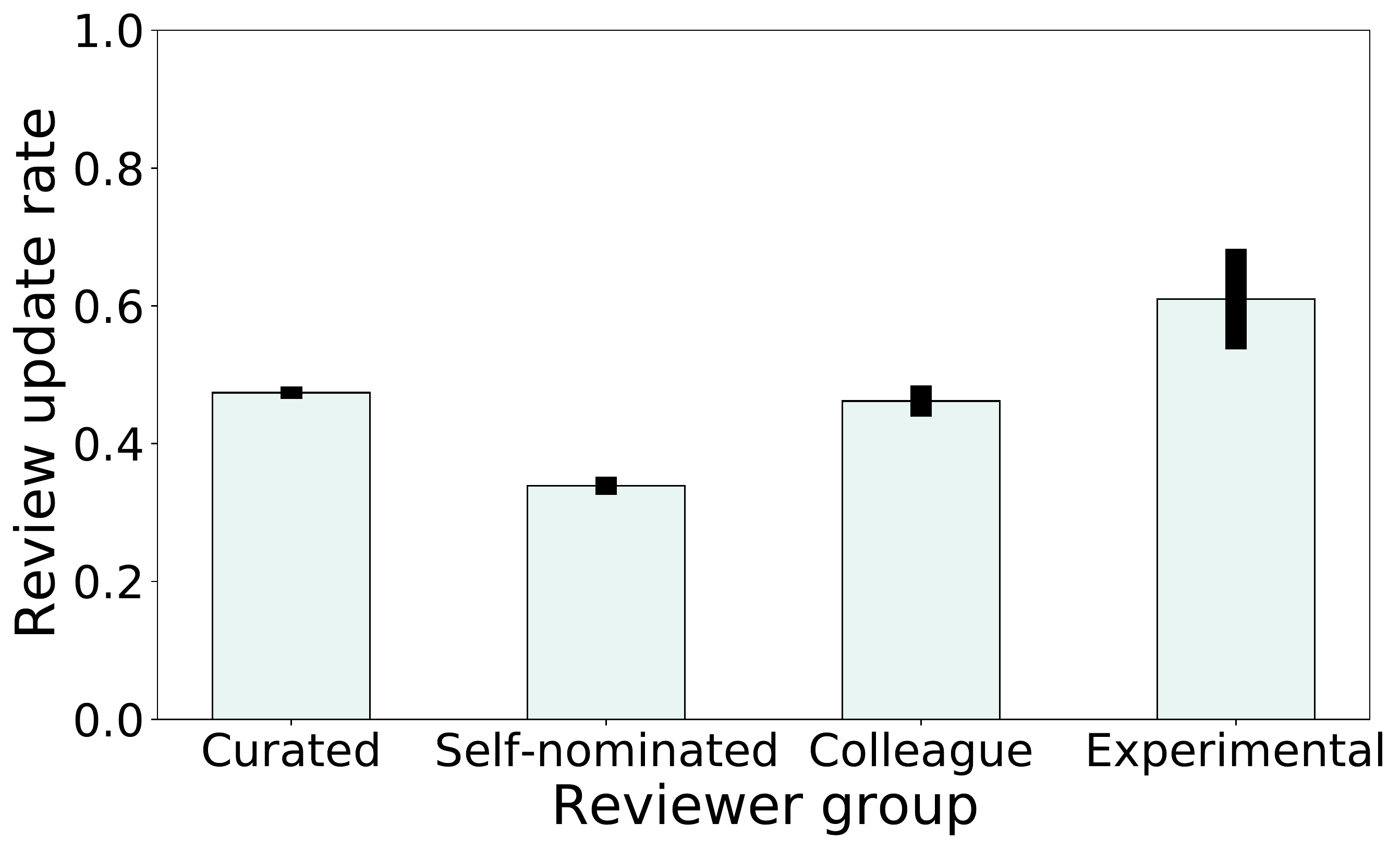}
        \caption{Post-rebuttal review update.}
    \label{fig:discussion:upd}
\end{subfigure}
\caption{Activity in the last stage of the review process.  {\sc Experimental} reviewers participate in the discussion and update textual reviews more often than other reviewers.}
\label{fig:discussion}
\end{figure}

Figure~\ref{fig:discussion} compares the fractions of paper-reviewer pairs such that the reviewer posted at least one message in the discussion thread (discussion activity rate, Figure~\ref{fig:discussion:disc}) / updated the textual review after the rebuttal (review update rate, Figure~\ref{fig:discussion:upd}), formal results of comparison are summarized in Table~\ref{table:discussion}. We note that in both dimensions \exprev{} reviewers are more active than other categories of reviewers.\footnote{Interestingly,  Figure~\ref{fig:discussion} shows that \snrev{} reviewers are less engaged in the last stage of the review process than senior \currev{} reviewers. This observation suggests that the relative engagement of junior \snrev{} reviewers decreases as the review process progresses. We do not see this in the \exprev{} reviewers and hypothesize that more tailored mentoring leads to a consistent engagement of \exprev{} reviewers.} 

\begin{table*}[ht]
\begin{center}
\begin{small}
\begin{sc}
\begin{tabular}{llccccr}
\toprule
  Criteria &  & Experimental & Main Pool & Curated & Self-Nominated & Colleague \\
\midrule
 & Sample Size & $123$ & $11985$ & $8298$ & $3687$ & $1262$ \\
 Discussion                              & Mean Value  & $0.68$ & $0.58$ & $0.60$ & $0.54$ & $0.60$ \\
 Activity                              & 90\% CI  & $[0.61; 0.75]$ & $[0.58; 0.59]$ & $[0.59; 0.61]$ & $[0.53; 0.56]$ & $[0.57; 0.62]$ \\
                               & $P$ value  & -- & $.033$ & $.083$ & $.004$ & $.077$ \\

\midrule
 & Sample Size & $123$ & $11985$ & $8298$ & $3687$ & $1262$ \\
{Review}                               & Mean Value & $0.61$ & $0.43$ & $0.47$ & $0.34$ & $0.46$ \\
{Update}     & 90\% CI & $[0.54; 0.68]$ & $[0.42; 0.44]$ & $[0.46; 0.48]$ & $[0.33; 0.35]$ & $[0.44; 0.48]$ \\
              & $P$ value  & -- & $<.001$ & $.005$ & $<.001$  & $.003$ \\

\bottomrule
\end{tabular}
\end{sc}
\end{small}
\end{center}
\caption{Comparison of post-rebuttal reviewer activity: participation in discussion (first 4 rows) and review update rate (last 4 rows). $P$ values are for the test of the difference of means between \exprev{} and each of the other groups of reviewers.} 
\label{table:discussion}
\end{table*}

\smallskip

\noindent \textbf{Review quality (Row 9 of Table~\ref{table:joint})} So far we have observed that \exprev{} reviewers are more active in all stages of the review process than reviewers from the main pool. However, the comparisons above do not decisively answer the question of \emph{quality} of reviews written by the new reviewers. To bridge this gap, we now report the evaluations of review quality made by meta-reviewers. At the end of the review process, meta-reviewers were asked to evaluate the quality of each review on a 3-point Likert item with the following options: ``Failed to meet expectations'' (Score 1), ``Met expectations'' (Score 2), ``Exceeded expectations'' (Score 3). Not all meta-reviewers completed the evaluation (approximately 25\% of all paper-reviewer pairs did not have a meta-reviewer evaluation) so for comparison of review quality we limit the attention to paper-reviewer pairs for which the corresponding meta-reviewer rated the quality of the review. Importantly, meta-reviewers were not aware of the group affiliation of reviewers. In the corresponding row of Table~\ref{table:joint}, we compare mean quality scores between different categories of reviewers and Table~\ref{table:rating} complements the comparison by additionally presenting a breakdown by reviewer groups, confirming that \exprev{} reviewers receive higher scores than their counterparts.

\begin{table}[b]
\begin{center}
\begin{small}
\begin{sc}
\begin{tabular}{lccccr}
\toprule
   & Experimental & Main Pool & Curated & Self-Nominated & Colleague \\
\midrule
Sample Size & $111$ & $11624$ & $8035$ & $3589$ & $1179$ \\
                               Mean Value & $2.26$ & $2.08$ & $2.11$ & $2.02$ & $2.11$ \\
     90\% CI & $[2.18; 2.34]$ & $[2.07; 2.09]$ & $[2.10; 2.12]$ & $[2.00; 2.03]$ & $[2.09; 2.14]$ \\
                               $P$ value  & -- & $<.001$ & $.002$ &  $<.001$ & $.003$ \\
\bottomrule
\end{tabular}
\end{sc}
\end{small}
\end{center}
\caption{Comparison of mean review qualities as evaluated by meta-reviewers. $P$ values are for the test of the difference of means between \exprev{} and each of the other groups of reviewers.} 
\label{table:rating}
\end{table}

\begin{figure}[h]
    \centering
    \includegraphics[width=8cm]{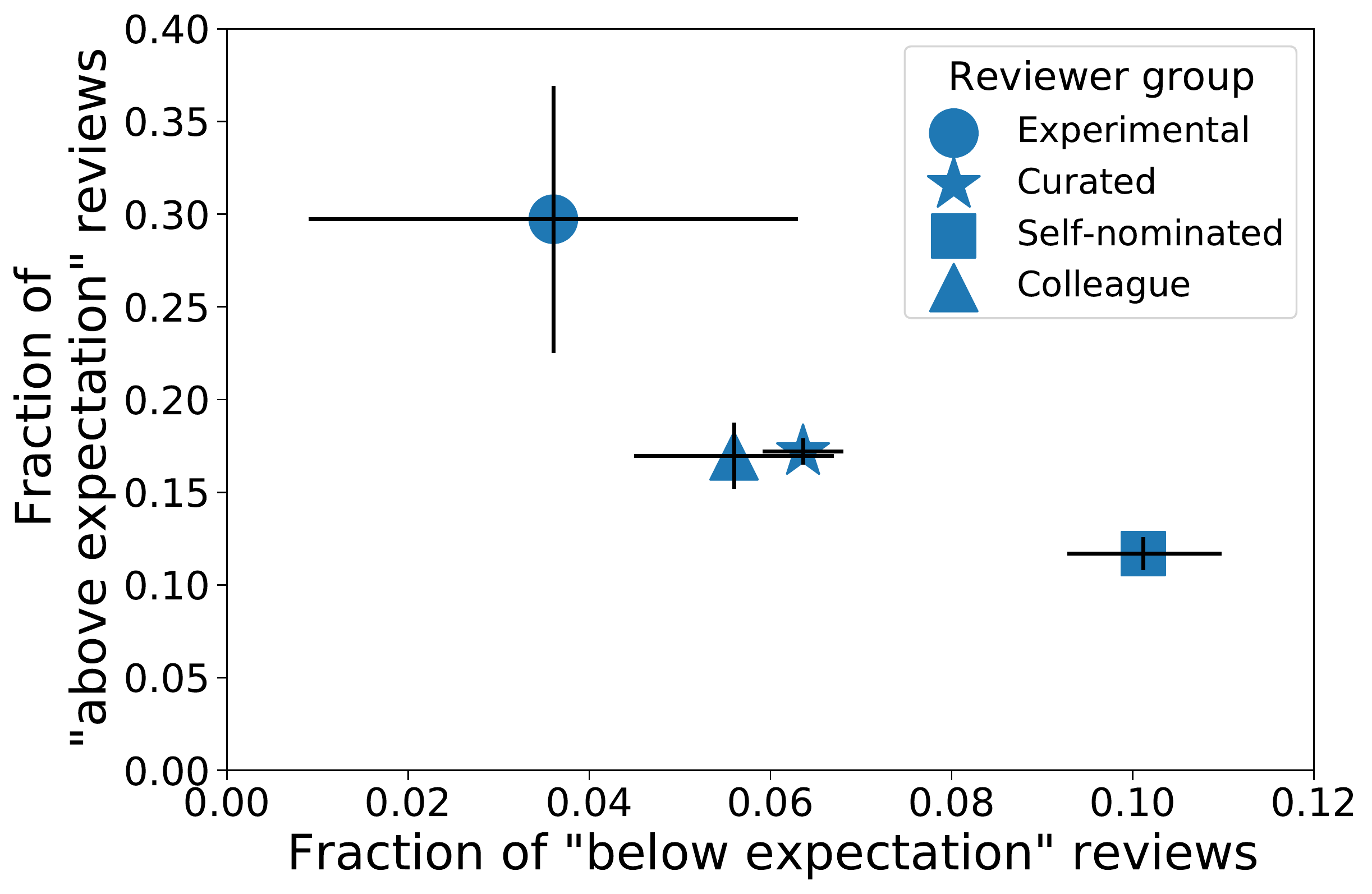}
    \caption{Evaluation of review quality by meta-reviewers. The closer the point to the upper-left corner, the better. {\sc Experimental} reviewers dominate other groups of reviewers.}
    \label{fig:rating}
\end{figure}

In contrasts to Tables~\ref{table:joint} and~\ref{table:rating} that compare mean ratings, Figure~\ref{fig:rating} visualizes the fraction of reviews below and above the expectations of meta-reviewers within each group. Observe that reviews written by \exprev{} reviewers exceed expectations of meta-reviewers more often than reviews of \currev{} and \snrev{} reviewers, and conditioning on the affiliation does not impact the comparison. Figure~\ref{fig:rating} also shows that \exprev{} reviewers produced substandard reviews less often than other reviewers.

\section{Discussion}
\label{section:discussion}

In this work we designed and executed an experimental procedure for novice reviewer recruiting and mentoring with a goal to address scarcity of qualified reviewers in large conferences. We evaluated the results of the experiment by juxtaposing the performance of the new reviewers to the traditional reviewer pool in the real \ICML{} 2020 conference. We now provide additional discussion of the recruiting and evaluation procedures, comment on the scalability of the experiment, and suggest directions for future work.

\subsection{Discussion of the Recruiting Experiment} 
\label{section:discussion:p1}

We begin from some important aspects of the auxiliary peer-review process we conducted to recruit reviewers. First, we perform the analysis of the demography of participants. After that, we mention another potentially useful aspect of the experiment related to reviews written in this auxiliary review process.

\smallskip

\noindent \textbf{Demography of participants} Recall that self-nominated individuals had to pass a publication and reviewing filters (see Section~\ref{section:meva} for details) to join the \snrev{} reviewer pool of \ICML{} 2020. We now test whether the subjects of our experiment satisfy these criteria. Table~\ref{table:demography} comprises relevant demographic information for 134 subjects of the selection experiment who completed the participation (that is, submitted the review of the assigned paper) and for 52 subjects who were eventually invited to join the \ICML{} 2020 \exprev{}  reviewer pool. Importantly, this demographic information was hidden from evaluators who performed the selection. Observe that most of the participants of our experiment (including those that were selected to join the \ICML{} reviewer pool) do not pass at least one of the filters mandatory for the \snrev{} reviewers. Similarly, none of the participants was invited to join the \currev{} reviewer pool. Therefore, we conclude that most of the participants of our experiment would not have been invited to \ICMLyear{} through conventional ways of reviewer recruiting.

\begin{table}[ht]
\begin{center}
\begin{small}
\begin{sc}
\begin{tabular}{lcc}
\toprule
          & All & Invited \\
\midrule
Total number 	& 134 & 52 	 \\ 
\midrule
With prior review experience & 36\% & 37\% \\ 
With publications & 72\% & 77\% \\ 
Pass reviewing filter & 21\% & 21\% \\ 
Pass publication filter & 24\% & 23\% \\ 
\textbf{Pass self-nominated filters}& \textbf{13\%} & \textbf{12\%}  \\ 
\bottomrule
\end{tabular}
\end{sc}
\end{small}
\end{center}
\caption{Demography of subjects of our experiment.}
\label{table:demography}
\end{table}

\smallskip

\noindent \textbf{Reviews} As a byproduct of our experiment, authors of manuscripts we used in the auxiliary peer-review process received a set of reviews. They generally admitted a high quality of reviews: several authors mentioned that reviewers found some errors/important typos in their papers or suggested some ways to improve the presentation. An author of one paper comments: 
\begin{quote}
    ``\emph{Very high quality reviews, in my opinion. Most of them [...] are clearly more detailed and more useful than reviews we received at [another top ML venue]}''
\end{quote}
and an author of another manuscript says:
\begin{quote}
    ``\emph{Those [reviews] were super informative and helpful!}''
\end{quote}
As a minor qualification, we underscore that while the overall feedback was positive, authors also helped us to identify a number of reviews of substandard quality for \ICML{} (with serious factual errors or dismissive criticism) and we found the author feedback to be very useful for the selection. That said, the positive overall feedback from authors hints that a large scale version of our experiment can give researchers a set of useful reviews before they submit a paper to a real conference, potentially decreasing the load on the actual conferences.

\subsection{Discussion of the Evaluation} 

We now discuss some aspects important for interpretation of the results of the comparison between \exprev{} reviewers and the main reviewer pool of \ICML{} 2020.

\paragraph{Aspect 1. Assignment procedure} Each paper submitted to the \ICML{} conference was first automatically assigned to three reviewers from the main pool, two of whom were \currev{} reviewers and one was a \snrev{} reviewer. In that, we tried to satisfy reviewer bids and optimize for the notion of textual similarity~\citep{charlin13tpms} between submissions and assigned reviewers, subject to a requirement that each reviewer is assigned at most six papers (a small set of reviewers requested a lower quota) and under a fairness constraint that aims at balancing assignment quality across submissions~\citep{stelmakh2018pr4a}. After that, each \exprev{} reviewer was manually assigned to three submissions as a fourth reviewer. All assignments were finally adjusted by meta-reviewers before being released to the reviewers. Given the small number of \exprev{} reviewers and the large number of submissions, the constraints we had to satisfy in the manual assignment were mild as compared to the main assignment. Hence, \exprev{} reviewers could receive submissions that better fit their expertise than reviewers from the main pool. 

Table~\ref{table:assignment} compares several metrics of assignment quality across reviewers from the main pool and \exprev{} reviewers. First, we note that \exprev{} reviewers were intentionally assigned fewer papers than reviewers from the main pool to ensure a gentle introduction to the review process. We underscore that in this study we aim to show that reviewers from the population not covered by current recruiting methods can usefully augment the reviewer pool given some special treatment (careful recruiting, reduced load and mentoring). Hence, we do not consider the difference in load to be a confounder in our comparisons. 

\begin{table}[t]
\begin{center}
\begin{small}
\begin{sc}
\begin{tabular}{lcc}
\toprule
          & Main Pool  & Experimental  \\
\midrule
\# Reviewers & 3012 & 52 \\
Mean reviewer load & 5.4 & 3.0  \\
Fraction of positive bids  & $0.88$ & $0.99$ \\
Mean paper$\leftrightarrow$reviewer similarity & $0.77$ & $0.88$\\
\bottomrule
\end{tabular}
\end{sc}
\end{small}
\end{center}
\caption{Assignment quality. ``Fraction of positive bids'' represents a fraction of paper-reviewer pairs in the assignment such that the reviewer has bid positively on the paper. Similarities between papers and reviewers take values in the interval $[0, 1]$ and are computed using the TPMS system~\citep{charlin13tpms}. 706 reviewers in the main pool and 4 \exprev{} reviewers did not have similarities computed and are excluded from the computation of mean similarity.} 
\label{table:assignment}
\end{table}

Second, \exprev{} reviewers were assigned to papers they positively bid on and to papers with high textual similarity more often than reviewers from the main pool. Hence, we caveat that the quality of the assignment differs significantly between reviewers from the main pool and \exprev{} reviewers, introducing a confounding factor that can potentially impact the comparison. On the other hand, the difference in the assignment quality may in part be due to the difference in bidding activity that we outlined in Section~\ref{section:evaluation}: a large number of positive bids made by \exprev{} reviewers gives more flexibility in satisfying them, thereby increasing the quality of the assignment. It will be of interest to investigate, in any future larger scale studies, whether a larger number of \exprev{} reviewers also continue to have such a higher quality of assignment due to higher bidding activity, and if not, then it will be of interest to observe how that impacts the other metrics.

\paragraph{Aspect 2. Quality evaluation} Following a standard approach in AI/ML conferences that conduct a survey of meta-reviewers on the review quality, when asking meta-reviewers to evaluate the quality of reviews, we left it for the meta-reviewers to decide on their expectations and did not precisely define the term ``quality''. As a result, different meta-reviewers could have different standards in mind, leading to some inconsistency in evaluations. To account for this issue, in the Appendix we complement the above analysis of review quality and some other metrics with additional comparisons performed on a restricted set of submissions that had at least one \exprev{} reviewer assigned (by doing so we equalize the sets of meta-reviewers who rate \exprev{} reviewers and other categories of reviewers as well as other characteristics of papers assigned to different groups of reviewers). Importantly, this analysis leads to the same conclusions as the analysis we described in Section~\ref{section:evaluation}.

Another related caveat is that the absence of a well-defined notion of quality could result in the substitution bias in meta-reviewers' judgments of review quality. For example, meta-reviewers' evaluations could be driven by the length of the review or some other computationally inexpensive, but suboptimal, proxy, resulting in a biased evaluation of quality. We urge the reader to be aware of this issue when interpreting the results of the quality evaluations.

\paragraph{Aspect 3. Insights from NeurIPS 2020 review process} After the experiment we describe in this paper was completed, the NeurIPS 2020 conference released the analysis of its review process~\citep{lin20data} in which they compared the quality of reviews written by curated and author-sourced reviewers (to avoid ambiguity, in this section we refer to curated NeurIPS reviewers as invited reviewers). In terms of the selection criteria, these groups of reviewers roughly correspond to \currev{} and \snrev{} groups we consider in the present paper, with an important exception that in our case \snrev{} reviewers were not required to have their paper submitted to \ICMLyear{}. There may also be some subtle differences in how the pools of invited NeurIPS reviewers and \currev{} \ICML{} reviewers were constructed. With these caveats, the analysis of NeurIPS 2020 data presents two key insights relevant to our study that we now discuss.

First, NeurIPS data suggests that the quality (as measured by meta-reviewers) of reviews written by author-sourced reviewers is only marginally worse than that of invited reviewers. This observation agrees with what we report in Table~\ref{table:rating} where \currev{} reviewers have slightly higher mean review quality than their \snrev{} counterparts, but the difference is somewhat more pronounced in our case.

Second, and perhaps more importantly, it appears that in NeurIPS the quality of reviews was negatively correlated with experience of reviewers: reviewers for whom NeurIPS 2020 was the first big ML conference they serve for appear to produce reviews of higher quality than their counterparts who have reviewed for major conferences before. In ICML 2020, all but perhaps 5--10\% of members of the main reviewer pool had past experience of being a reviewer for some top ML venues,\footnote{Some \currev{} reviewers were recommended by the meta-reviewers and are not guaranteed to have the past review experience} while most of \exprev{} reviewers did not have such experience. Hence, the result of NeurIPS analysis highlights another potential confounding factor in our study: past reviewing experience. Indeed, hypothetically the results of our experiment can be explained by the fact that reviewers put more efforts into their first reviews and then become less engaged in future conferences. 

Let us now qualify the above caveat. First, in our experiment the difference between \exprev{} reviewers and reviewers from the main \ICML{} reviewer pool appears to be considerably larger than the difference between the first-time reviewers and those who have reviewed before observed in NeurIPS. The larger effect size hints at the potential effect of our intervention. Second, in the present work we carefully account for the affiliation confounding factor and additionally juxtapose the \exprev{} reviewers with the \elrev{} reviewers who share the same affiliation. The NeurIPS analysis does not provide such comparison and hence we cannot remove this confounding factor. Finally, in the present paper we compare the performance of reviewers on various metrics beyond the review quality and it would be interesting to see how novice NeurIPS reviewers perform on these metrics.

\paragraph{Aspect 4. Additional caveats} In addition to the caveats mentioned above, we would like to make several other remarks:

\begin{itemize}[itemsep=2pt, leftmargin=*]
    \item First, while we measured a number of metrics that were possible to measure and have been considered in the literature,  we cannot exclude the possibility that \exprev{} reviewers may be worse than the main pool of reviewers in some other aspect  not considered here.
    
    \item  Second, it is possible that behavior of \exprev{} reviewers was affected by demand characteristics~\cite{mccambridge2012effects}, that is, \exprev{} reviewers could hypothesize that we want them to perform better than reviewers from the main pool and hence they could adjust their behaviour to meet these perceived expectations.
    
    \item Third, the review process of the \ICMLyear{} conference was impacted by the COVID-19 pandemic and the impact of the pandemic on different reviewer groups could be unequal. For example, senior reviewers from the \currev{} pool could have more family-related duties (and hence could be more restricted in reviewing ability) than junior \exprev{} reviewers.
    
    \item Finally, in extrapolating any results to other conferences, one should carefully consider any idiosyncrasies of specific conferences.     
\end{itemize}
All the aforementioned caveats coupled with sensitivity of the subject matter underscore the importance of a careful experimentation with the proposed procedure before its implementation in the routine review process.

\smallskip

\noindent \textbf{Aspect 5. The role of reviewers} With the above caveats, the experiment demonstrated that \exprev{} reviewers are comparable to and sometimes even better than reviewers recruited in conventional ways in terms of various metrics analyzed in Section~\ref{section:evaluation}. However, we qualify that this observation absolutely does not imply that \exprev{} reviewers can entirely substitute the pool of experienced reviewers. Instead, we conclude that if recruited and mentored appropriately, \exprev{} reviewers can form a useful augmentation to the traditional pool. The \exprev{} and experienced reviewers may have different strengths that can be combined to achieve an overall improvement of the peer-review quality. For instance~\citep{blog2019decker}, \exprev{} and, more generally, junior reviewers can be used to evaluate nuanced technical details of submissions (e.g., proofs) while senior researchers can focus on the broader picture and more subjective criteria such as impact where their expertise is extremely important.

\subsection{Scalability of the Experiment}

In this section we provide some ideas on how the experiment we described in this paper can be scaled to increase the number of recruited reviewers from 52 to several hundred. To this end, recall that the experiment is based on the two major components: the selection and mentoring mechanisms. We now comment on how to scale each of these components.

\medskip

\noindent \textbf{Selection Mechanism} The current selection pipeline requires an amount of work equivalent to 2 to 4 days of the conference workflow chair's work and 2 to 4 hours of the conference program chairs' work to execute the experiment. Hence, it is important to design a version of the selection mechanism that can handle more reviewers while not resulting in a proportional increase of the load on the organizers. 

First, we note that a large share of work in the selection stage was spent on finding papers to use in the auxiliary review. For this initial experiment, we had not made the call for papers public and instead personally reached to dozens of colleagues asking them to contribute their manuscripts which resulted in a large amount of communication-related work. However, this initial experiment demonstrated a lot of enthusiasm from authors of the papers used in the experiment who appreciated  a set of useful reviews they received. Hence, we believe that we can easily extend the pool of papers by widely distributing the call for papers.

Similarly, participants of the selection mechanism executed in the present study were active in signing up for the experiment and appreciated an opportunity to join the \ICML{} reviewer pool. In this initial experiment, the population of participants was limited to students of 5 large US universities. Hence, by making an open call for participants on behalf of a large ML conference, we expect to increase the pool of candidates to several hundred participants without much additional efforts.

The major part of the selection mechanism that requires a close attention of organizers is the review evaluation and final decision making. As noted in Section~\ref{section:mech}, we found author feedback to be very helpful for evaluating the quality of reviews and hence the selection part can be streamlined if the authors are required to evaluate all the reviews received in the experiment. Given that authors of papers used in the experiment generally found these reviews useful, we think that such a requirement is feasible as these reviews serve as a good incentive for authors to put some efforts in the experiment. 

\medskip

\noindent \textbf{Mentoring} As mentioned in Section~\ref{section:mentoring}, the total amount of time and effort in the mentorship of 52 \exprev{} reviewers was equal to about half the time and effort for a meta-reviewer's job. We note that the time demand for mentorship does not increase proportionally to the number of reviewers as some parts of the mentorship have a fixed cost (e.g., sending general guideline emails or multiple reminders to non-responsive reviewers). Thus, several additional committee members recruited specifically for mentoring will allow to handle several hundred novice reviewers in the real conference. Alternatively, mentoring can be distributed across many senior researchers who do not have time for meta-reviewing, but can contribute a smaller amount to mentoring.

\medskip

Overall,  we believe that the procedure outlined above can produce several hundred experimental reviewers without overburdening the organizers of the experiment.

\subsection{Future Work}

An important direction for future work is to compare \exprev{} reviewers with the main reviewer pool in a larger scale study whose design we outlined above. A larger experiment would enable a deeper analysis of textual reviews written by different reviewer groups. It will also be of interest to design and execute experiments that address the caveats discussed above.

Another important direction is a principled design of a mentoring protocol to support novice reviewers. Since ML/AI conferences have hundreds of meta-reviewers, it may be prudent to assign a small number of meta-reviewers as mentors for junior reviewers and reduce their meta-reviewer workload accordingly. Future editions could also involve sharing more material on how to review with reviewers (e.g.,~\citealp{kohler2020supporting}) and holding webinars with Q\&A sessions. 

Finally, the feedback from the \exprev{} reviewers was that it was helpful for them to experience and gain insights into the review process, which will also help in their own research dissemination in the future. It would be interesting to measure the impact of the guided introduction to the review process in the early stages of career on the future trajectory of the individuals as researchers and reviewers.

\section*{Acknowledgments}

We thank authors of papers used in the selection experiment for contributing their works to this study and for helping us with identifying strong reviews. We also thank Devendra Singh Chaplot who served as a domain expert in the external evaluation of reviews. We are grateful to the support team of the Microsoft Conference Management Toolkit (CMT) who hosted our selection experiment and helped with many customization requests during the experiment and the actual \ICMLyear{} conference. Finally, we appreciate the efforts of all reviewers and meta-reviewers involved in the \ICMLyear{} review process. This study was approved by Carnegie Mellon University Institutional Review Board.

This work was supported in part by NSF CAREER award 1942124 and in part by NSF CIF 1763734. 

{\small 
\bibliography{references.bib}
}
\bibliographystyle{apalike}

\vspace{1cm}

\appendix

\noindent \textbf{\Large{Appendix: Additional Evaluation Details}}\\


\noindent In this section we provide additional details for comparison of \exprev{} reviewers with reviewers from the main reviewer pool of the \ICML{} 2020 conference. Specifically, where applicable we replicate comparisons described in the main paper, conditioning on a target set of papers with at least one \exprev{} reviewer assigned. Note that this conditioning significantly reduces the sample size and hence we may not always have enough data to establish statistically significant differences. Nevertheless, the results we present below allow to make some qualitative conclusions. 

Before we proceed, recall that the main reviewer pool consists of complementary sets of \currev{} and \snrev{} reviewers and we additionally consider a subset of \elrev{} reviewers who are affiliated with one of the 5 US schools we were recruiting \exprev{} reviewers from. 

\begin{figure}[h]
    \centering
    \includegraphics[width=8cm]{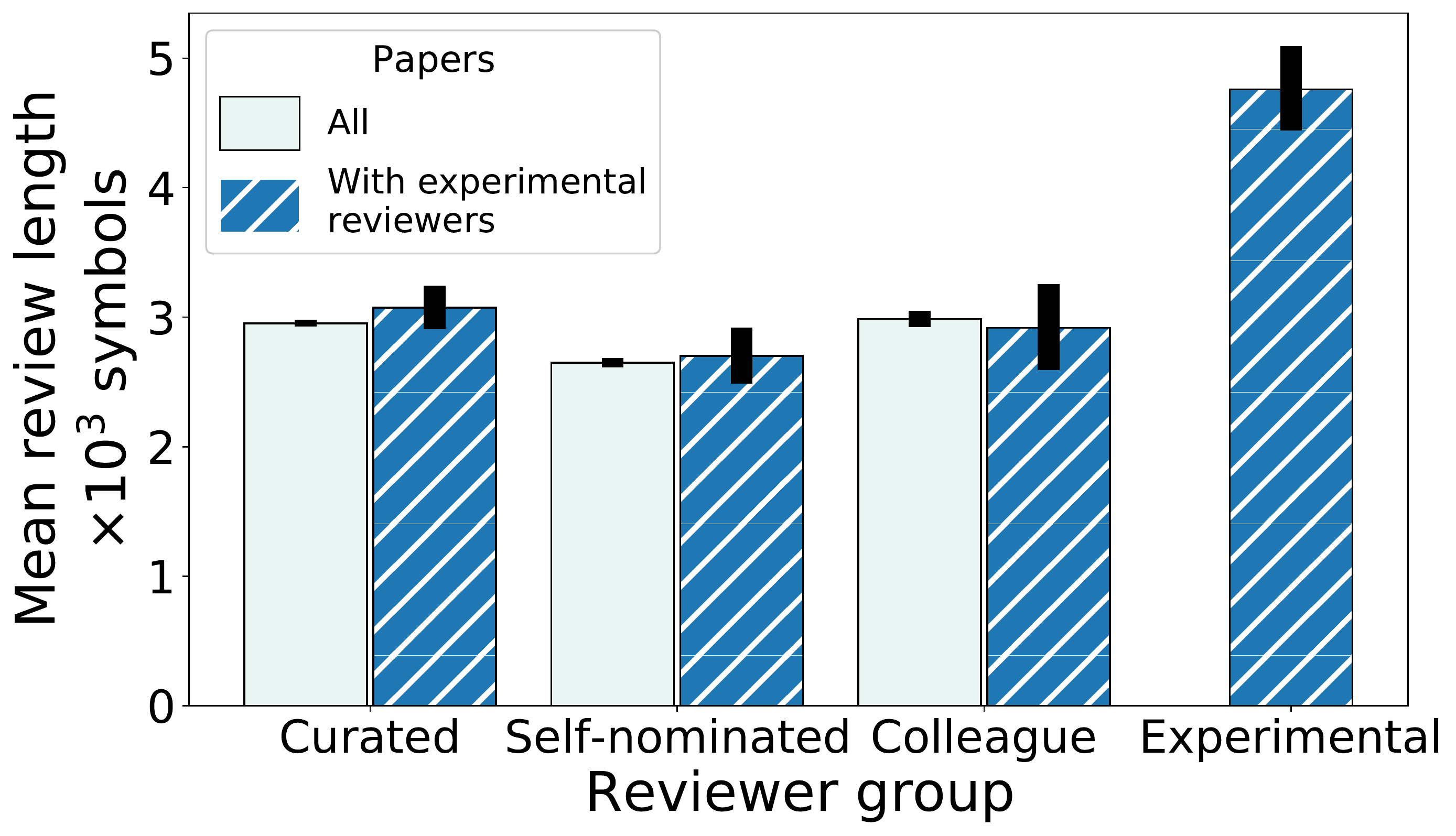}
    \caption{Mean review lengths (in symbols). {\sc Experimental} reviewers write longer reviews than other reviewers.}
    \label{fig:len}
\end{figure}

\paragraph{Review length (Row 3 of Table~\ref{table:joint})} Bidding activity as well as in-time review submission are independent of a set of papers assigned to reviewers so we begin our additional analysis from a comparison of mean review lengths. Figure~\ref{fig:len} compares mean lengths of initial reviews across different reviewer groups, where mean values are computed using all reviews and also using reviews written for papers that have at least one \exprev{} reviewer assigned. Overall, we observe that \exprev{} reviewers write considerably longer reviews than other reviewers even after conditioning on the aforementioned subset of paper. Table~\ref{table:len2} mimics Table~\ref{table:len} with the exception that only papers with \exprev{} reviewers assigned are used for the comparison. 

\begin{table*}[h]
\begin{center}
\begin{small}
\begin{sc}
\begin{tabular}{lccccr}
\toprule
 & Experimental & Main Pool & Curated & Self-Nominated & Colleague \\
\midrule
Sample Size & $154$ & $423$ & $294$ & $129$ & $51$ \\
Mean Value  & $4759$ & $2959$ & $3073$ & $2700$ & $2917$ \\
90\% CI  & $[4432; 5089]$ & $[2827; 3098]$ & $[2908; 3248]$ & $[2487; 2928]$ & $[2592; 3261]$\\
$P$ value & -- & $<.001$ & $<.001$ & $<.001$ & $<.001$ \\
\bottomrule
\end{tabular}
\end{sc}
\end{small}
\end{center}
\caption{Comparison of mean lengths (in symbols) of reviews using papers with at least one \exprev{} reviewer assigned. $P$ values are for the test of the difference of means between \exprev{} and each of the other groups of reviewers.} 
\label{table:len2}
\end{table*}


\paragraph{Hypercriticality (Row 4 of Table~\ref{table:joint})}

Analysis presented in the main paper did not reveal hypercriticality in \exprev{} reviewers. Table~\ref{table:hyper2} replicates the analysis displayed in Table~\ref{table:hyper} on the target subset of papers and also does not show any evidence of hypercriticality in \exprev{} reviewers. 

\begin{table*}[h]
\begin{center}
\begin{small}
\begin{sc}
\begin{tabular}{lccccr}
\toprule
   & Experimental & Main Pool & Curated & Self-Nominated & Colleague \\
\midrule
Sample Size & $154$ & $423$ & $294$ & $129$ & $51$ \\
Mean Value  & $3.34$ & $3.22$ & $3.20$ & $3.27$ & $3.12$ \\
90\% CI  & $[3.17; 3.51]$ & $[3.13; 3.32]$ & $[3.10; 3.32]$ & $[3.09; 3.46]$ & $[2.86; 3.37]$\\
$P$ value  & -- & $.361$ & $.282$ & $.691$ & $.325$ \\
\bottomrule
\end{tabular}
\end{sc}
\end{small}
\end{center}
\caption{Comparison of mean initial overall scores using papers with at least one \exprev{} reviewer assigned. $P$ values are for the test of the difference of means between \exprev{} and each of the other groups of reviewers.} 
\vspace{-15pt}
\label{table:hyper2}
\end{table*}


\paragraph{Expertise and confidence (Rows 5 and 6 of Table~\ref{table:joint})}

Figure~\ref{fig:exp_conf2} juxtaposes the mean self-assessed confidence and expertise of reviewers computed over all papers and over papers with at least one \exprev{} reviewer assigned. Observe that conditioning on the target subset of papers does not change the conclusions we made in the main paper (see Table~\ref{table:conf_and_exp2} for formal comparison).

{
\begin{figure}[h]
\centering
\begin{subfigure}[t]{0.48\textwidth}%
    \vskip 0pt
    \centering
    \includegraphics[width=7cm]{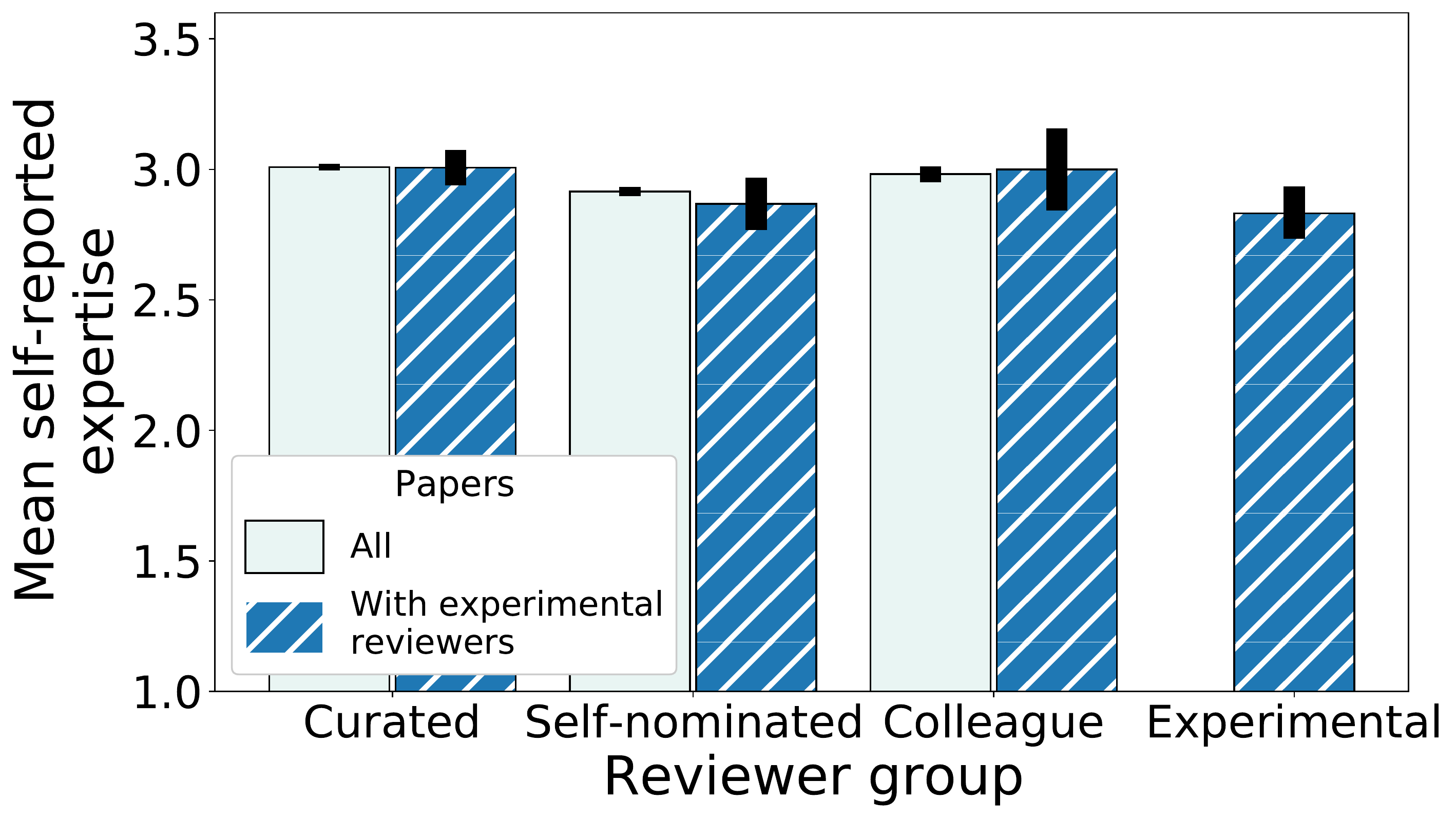}
    \caption{Mean self-reported expertise. {\sc Experimental} reviewers report considerably lower expertise than other groups of reviewers.}
    \label{fig:experience2}
\end{subfigure}\hspace{5pt}%
\begin{subfigure}[t]{0.48\textwidth}%
    \vskip 0pt
    \centering
    \includegraphics[width=7cm]{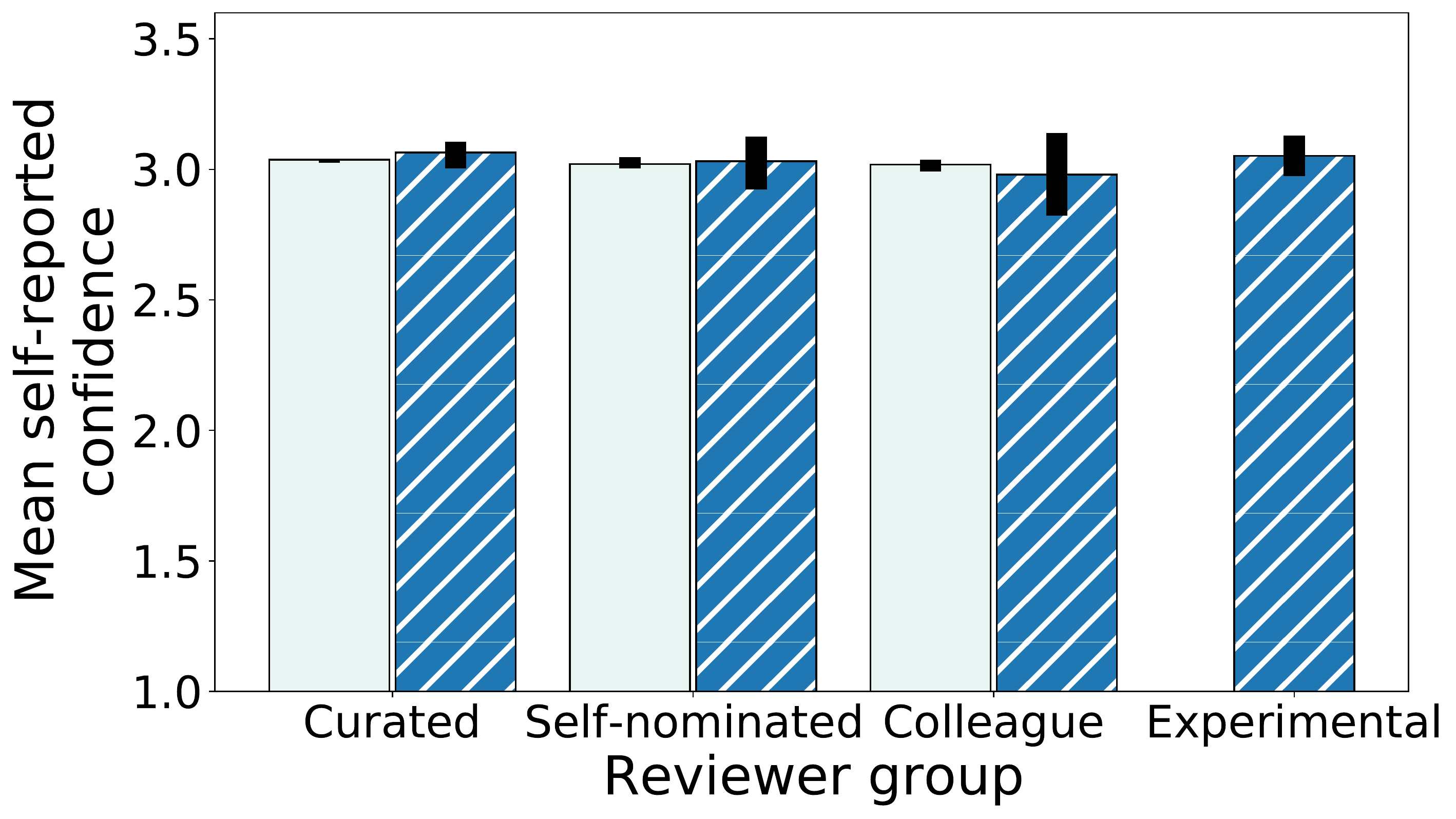}
    \caption{Mean self-reported confidence. We do not observe significant difference in mean confidence of different reviewer groups.}
    \label{fig:conf2}
\end{subfigure}
\caption{Comparison of self-assessed expertise and confidence.}
\vspace{-15pt}
\label{fig:exp_conf2}
\end{figure}
}

\begin{table*}[h!]
\begin{center}
\begin{small}
\begin{sc}
\begin{tabular}{llccccr}
\toprule
  Criteria &  & Experimental & Main Pool & Curated & Self-Nominated & Colleague \\
\midrule
\multirow{4}{*}{Expertise} & Sample Size & $154$ & $423$ & $294$ & $129$ & $51$ \\
                               & Mean Value & $2.83$ & $2.96$ & $3.01$ & $2.87$ & $3.00$ \\
     & 90\% CI & $[2.73; 2.94]$ & $[2.91; 3.02]$ & $[2.94; 3.07]$ & $[2.77; 2.97]$ & $[2.84; 3.16]$ \\
                               & $P$ value  & -- & $.064$ & $.021$ & $.737$ & $.211$ \\
\midrule
\multirow{4}{*}{Confidence} & Sample Size & $154$ & $423$ & $294$ & $129$ & $51$ \\
                               & Mean Value  & $3.05$ & $3.05$ & $3.06$ & $3.03$ & $2.98$ \\
                               & 90\% CI  & $[2.97; 3.13]$ & $[3.00; 3.11]$ & $[3.00; 3.13]$ & $[2.92; 3.14]$ & $[2.82; 3.14]$ \\
                               & $P$ value  & -- & $.993$ & $.878$ & $.863$ & $.556$ \\
\bottomrule
\end{tabular}
\end{sc}
\end{small}
\end{center}
\caption{Comparison of self-assessed expertise (first 4 rows)  and confidence (last 4 rows) using papers with at least one \exprev{} reviewer assigned.  $P$ values are for the test of the difference of means between \exprev{} and each of the other groups of reviewers.} 
\label{table:conf_and_exp2}
\end{table*}


\paragraph{Rebuttals and discussion (Rows 7 and 8 of Table~\ref{table:joint})}

\begin{figure}[t]
\centering
\begin{subfigure}[t]{0.48\textwidth}%
    \vskip 0pt
    \centering
    \includegraphics[width=7cm]{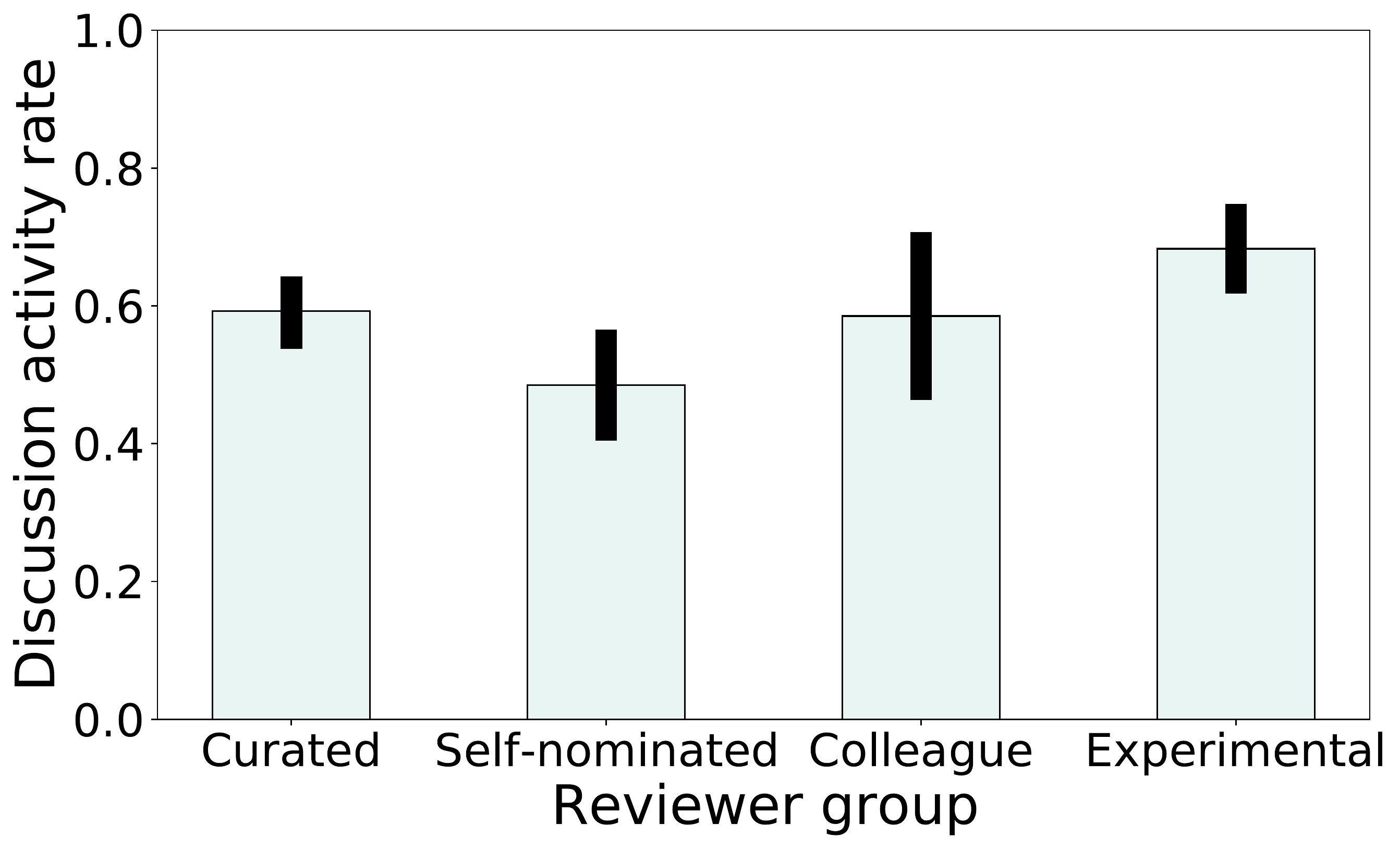}
    \caption{Participation in discussion.}
    \label{fig:discussion2:disc}
\end{subfigure}\hspace{5pt}%
\begin{subfigure}[t]{0.48\textwidth}%
    \vskip 0pt
    \centering
    \includegraphics[width=7cm]{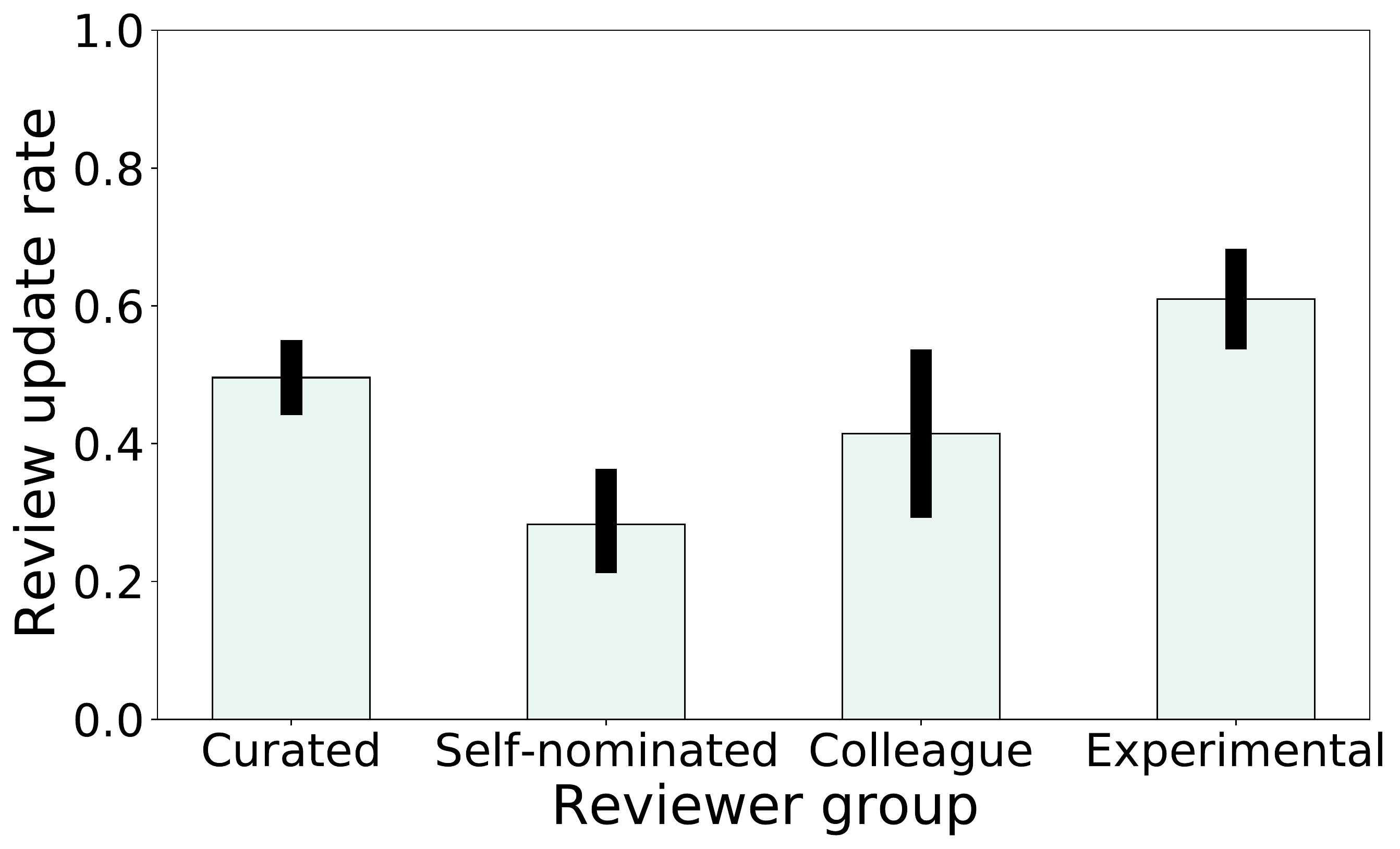}
        \caption{Post-rebuttal review update.}
    \label{fig:discussion2:upd}
\end{subfigure}
\caption{Activity of reviewers in the last stage of the review process conditioned on papers with at least one \exprev{} reviewer assigned.  {\sc Experimental} reviewers participate in the discussion and update reviews more actively than other reviewers.}
\label{fig:discussion2}
\end{figure}

We now provide additional details on comparison of reviewers' activity in the post-rebuttal stage of the conference peer-review process. Recall that for this comparison we use only paper-reviewer pairs such that the authors of the paper supplied the response to the initial reviews. Figure~\ref{fig:discussion2} replicates Figure~\ref{fig:discussion} with the exception that it is computed using papers that have at least one \exprev{} reviewer assigned. Note that even after conditioning on this subset of papers, \exprev{} reviewers remain to be more active in the post-rebuttal stage of the review process than other categories of reviewers. (see  Table~\ref{table:discussion2} for the formal comparison).

\begin{table*}[t]
\begin{center}
\begin{small}
\begin{sc}
\begin{tabular}{llccccr}
\toprule
  Criteria &  & Experimental & Main Pool & Curated & Self-Nominated & Colleague \\
\midrule
 & Sample Size & $123$ & $337$ & $238$ & $99$ & $41$ \\
 Discussion                              & Mean Value  & $0.68$ & $0.56$ & $0.59$ & $0.48$ & $0.59$ \\
 Activity                              & 90\% CI  & $[0.61; 0.75]$ & $[0.52; 0.61]$ & $[0.54; 0.64]$ & $[0.40; 0.57]$ & $[0.46; 0.71]$ \\
                               & $P$ value  & -- & $.026$ & $.119$ & $.003$ & $.338$ \\

\midrule
 & Sample Size & $123$ & $337$ & $238$ & $99$ & $41$ \\
{Review}                               & Mean Value & $0.61$ & $0.43$ & $0.50$ & $0.28$ & $0.41$ \\
{Update}     & 90\% CI & $[0.54; 0.68]$ & $[0.39; 0.48]$ & $[0.44; 0.55]$ & $[0.21; 0.35]$ & $[0.29; 0.54]$ \\
              & $P$ value  & -- & $.002$ & $.050$ & $<.001$ & $.045$ \\

\bottomrule
\end{tabular}
\end{sc}
\end{small}
\end{center}
\caption{Comparison of post-rebuttal reviewer activity using papers with at least one \exprev{} reviewer assigned: participation in discussion (first 4 rows) and review update rate (last 4 rows). $P$ values are for the test of the difference of means between \exprev{} and each of the other groups of reviewers.} 
\label{table:discussion2}
\end{table*}


\paragraph{Review quality (Row 9 of Table~\ref{table:joint})}

\begin{figure}[ht]
    \centering
    \includegraphics[width=9cm]{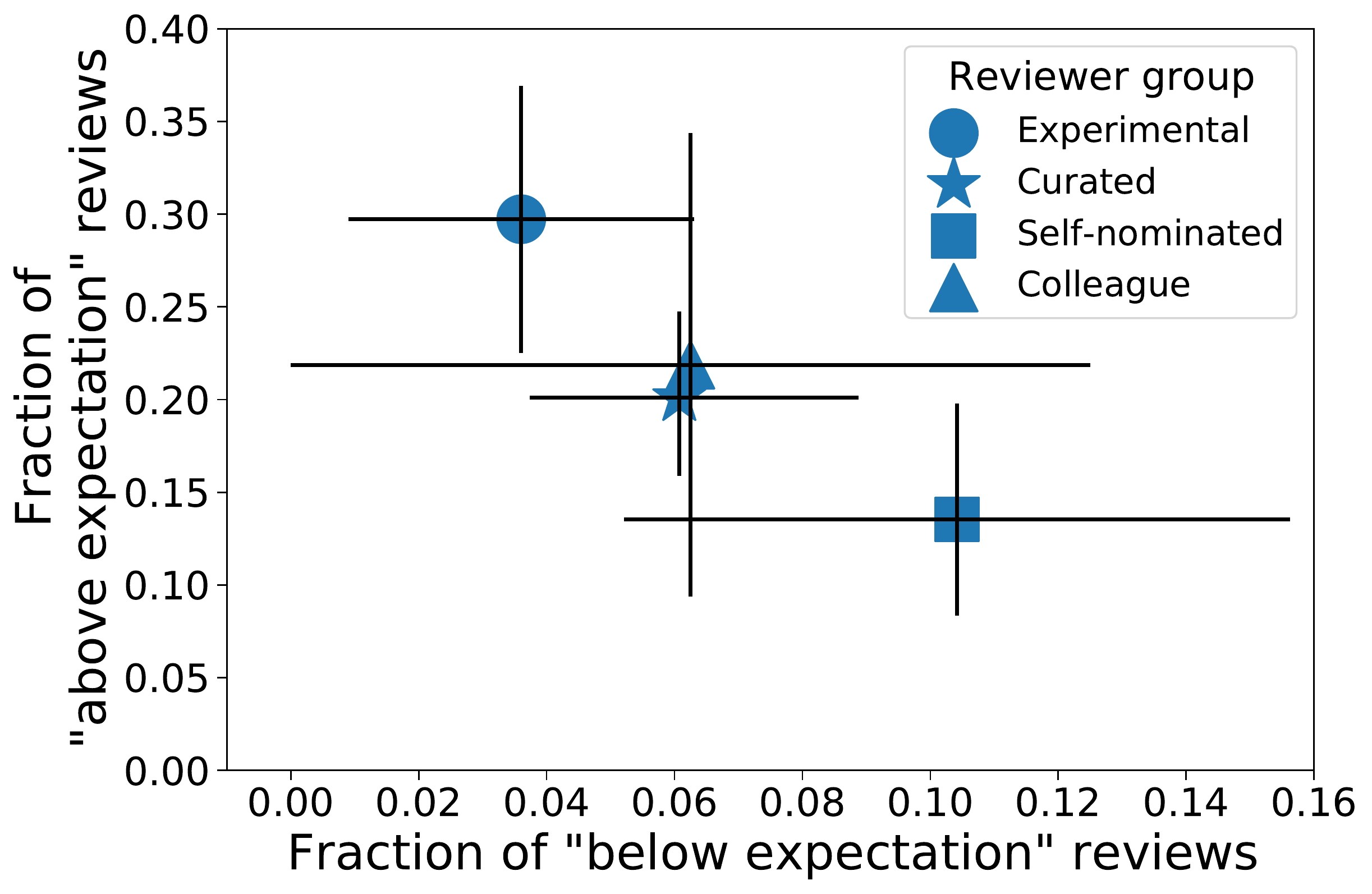}
    \caption{Evaluation of mean review quality by meta-reviewers conditioned on papers with at least one \exprev{} reviewer assigned. The closer the point to the upper-left corner, the better. {\sc Experimental} reviewers dominate other groups of reviewers.}
    \label{fig:rating2}
\end{figure}

We conclude the analysis with comparison of review quality evaluated by meta-reviewers. First, Figure~\ref{fig:rating2} replicates Figure~\ref{fig:rating} with the exception that we condition on papers that have at least one \exprev{} reviewer assigned. As before, conditioning on this subset of papers does not change the qualitative relationship between different groups of reviewers with \exprev{} pool dominating others. Getting back to the comparison of the mean quality scores that we reported in Tables~\ref{table:joint} and~\ref{table:rating}, we now complement this analysis with results reported in Table~\ref{table:rating2}. Again, we conclude that \exprev{} reviewer remain to have higher mean quality of reviews even after we equalize the sets of meta-reviewers who rate the reviews written by different groups of reviewers.

\begin{table*}[ht]
\begin{center}
\begin{small}
\begin{sc}
\begin{tabular}{lccccr}
\toprule
 & Experimental & Main Pool & Curated & Self-Nominated & Colleague \\
\midrule
Sample Size & $111$ & $310$ & $214$ & $96$ & $32$ \\
Mean Value  & $2.26$ & $2.11$ & $2.14$ & $2.03$ & $2.16$ \\
90\% CI  & $[2.18; 2.34]$ & $[2.06; 2.15]$ & $[2.08; 2.20]$ & $[1.95; 2.11]$ & $[2.00; 2.31]$ \\
$P$ value  & -- & $.008$ & $.058$ & $.002$ & $.431$ \\
\bottomrule
\end{tabular}
\end{sc}
\end{small}
\end{center}
\caption{Comparison of mean review qualities as evaluated by meta-reviewers using papers with at least one \exprev{} reviewer assigned. $P$ values are for the test of the difference of means between \exprev{} and each of the other groups of reviewers.} 
\label{table:rating2}
\end{table*}

\end{document}